%% file: manuscript.tex
\documentclass[]{acmart}
\usepackage{booktabs} 

\usepackage[export]{adjustbox}

\usepackage[ruled]{algorithm2e} 

\usepackage{array,multirow}

\usepackage{listings}
\usepackage{color}

\definecolor{codegreen}{rgb}{0,0.6,0}
\definecolor{codegray}{rgb}{0.5,0.5,0.5}
\definecolor{codepurple}{rgb}{0.58,0,0.82}
\definecolor{backcolour}{rgb}{1,1,1}

\lstdefinestyle{mystyle}{
	backgroundcolor=\color{backcolour},   
	commentstyle=\color{codegreen},
	keywordstyle=\bfseries\color{black},
	numberstyle=\tiny\color{codegray},
	stringstyle=\color{codepurple},
	basicstyle=\footnotesize,
	breakatwhitespace=false,         
	breaklines=true,                 
	captionpos=b,                    
	keepspaces=true,                 
	numbers=left,                    
	numbersep=5pt,                  
	showspaces=false,                
	showstringspaces=false,
	showtabs=false,                  
	tabsize=2
}

\setcopyright{none}

\usepackage[xindy]{glossaries} 
\input{glossary}

\usepackage{subcaption}

\begin{document}
	\title[Low Complexity MAC units for Convolutional Neural Networks]{Low Complexity Multiply-Accumulate Units for Convolutional Neural Networks with Weight-Sharing} 
	
	\author{James Garland}
        \author{David Gregg}
	\orcid{0000-0002-8688-9407}
	\affiliation{%
		\institution{Trinity College Dublin}
		\streetaddress{Westland Row}
		\city{Dublin}
		\state{Dublin}
		\postcode{D02 DP70}
		\country{Ireland}}

	\affiliation{%
		\institution{Trinity College Dublin}
		\city{Dublin}
		\country{Ireland}
	}
	
	\renewcommand\shortauthors{J. Garland and D. Gregg}
	
	\begin{abstract}
		\acrfullpl{cnn} are one of the most successful machine learning techniques for image, voice and video processing. \glspl{cnn} require large amounts of processing capacity and memory bandwidth. Hardware accelerators have been proposed for \glspl{cnn} which typically contain large numbers of \gls{mac} units, the multipliers of which are large in \gls{ic} gate count and power consumption. ``Weight sharing'' accelerators have been proposed where the full range of weight values in a trained \gls{cnn} are compressed and put into bins and the bin index used to access the weight-shared value. We reduce power and area of the \gls{cnn} by implementing \acrfull{pasm} in a weight-shared \gls{cnn}. \gls{pasm} re-architects the \gls{mac} to instead count the frequency of each weight and place it in a bin. The accumulated value is computed in a subsequent multiply phase, significantly reducing gate count and power consumption of the \gls{cnn}. In this paper, we implement \gls{pasm} in a weight-shared \gls{cnn} convolution hardware accelerator and analyze its effectiveness. Experiments show that for a clock speed 1GHz implemented on a 45nm \gls{asic} process our approach results in fewer gates, smaller logic, and reduced power with only a slight increase in latency. We also show that the same weight-shared-with-\gls{pasm} \gls{cnn} accelerator can be implemented in resource-constrained \glspl{fpga}, where the \gls{fpga} has limited numbers of \gls{dsp} units to accelerate the \gls{mac} operations.
	\end{abstract}
	
	%
	%

	%
	%
	
	
	\thanks{This research is supported by Science Foundation Ireland, Project 12/IA/1381. We also thank the Institute of Technology Carlow, Carlow, Ireland for their support.
		
	\textbf{Extended paper:} This paper is an expanded version of a short, four page paper that appeared in \acrshort{ieee} \gls{cal} (\cite{LowComplexityMacForWeightSharingCnn2017:Garland}). \acrshort{ieee} \gls{cal}'s publication policy is that ``due to the short format, we expect that publication in \acrshort{ieee} \gls{cal} should not preclude subsequent publication in top-quality conferences or full-length journals''. Our earlier \acrshort{ieee} \gls{cal} short paper proposed our \gls{pasm} unit for multiply-accumulate operations, and provided an evaluation of the unit in isolation using an \gls{asic} design flow. In contrast the current paper provides much greater detail and analysis, and evaluates our \gls{pasm} unit in the context of a \gls{cnn} hardware accelerator rather than in isolation. In the current paper we also study the effectiveness of \gls{pasm} for a \gls{cnn} accelerator on \glspl{fpga}, and provide experimental results.

	\textbf{Authors' addresses:}School of Computer Science and Statistics, Trinity College Dublin, Westland Row, Dublin, D02 DP70, Ireland, jgarland@tcd.ie, david.gregg@cs.tcd.ie.}
	
	\maketitle
	
	
	\input{body}

\end{document}

%% file: glossary.tex
\newacronym{cpu}{CPU}{central processing unit}
\newacronym{cnn}{CNN}{Convolutional neural network}
\newacronym{dnn}{DNN}{deep neural network}
\newacronym{mac}{MAC}{multiply-accumulate}
\newacronym{asic}{ASIC}{application specific integrated circuit}
\newacronym{ieee}{IEEE}{Institute of Electrical and Electronic Engineers}
\newacronym{cal}{CAL}{Computer Architecture Letters}
\newacronym{ic}{IC}{integrated circuit}
\newacronym{fpga}{FPGA}{field programmable gate array}
\newacronym{dsp}{DSP}{digital signal processor}
\newacronym{bram}{BRAM}{block RAM}
\newacronym{relu}{ReLU}{rectified linear unit}
\newacronym{gpu}{GPU}{graphics processor unit}
\newacronym{pas}{PAS}{parallel accumulate and store}
\newacronym{pasm}{PASM}{parallel accumulate shared MAC}
\newacronym{rtl}{RTL}{register transfer logic}
\newacronym{ip}{IP}{intellectual property}
\newacronym{hdl}{HDL}{hardware description language} 
\newacronym{sdc}{SDC}{Synopsys design constraint}
\newacronym{xdc}{XDC}{Xilinx design constraint}
\newacronym{iob}{IOB}{input output buffer}
\newacronym{ram}{RAM}{random access memory}
\newacronym{xpe}{XPE}{Xilinx power estimator}
\newacronym{clb}{CLB}{configurable logic block}
\newacronym{lut}{LUT}{look up table}
\newacronym{mux}{MUX}{multiplexer}
\newacronym{sop}{SOP}{sum-of-products}
\newacronym{mcmk}{MCMK}{multi-channel, multi-kernel convolution}
\newacronym{sram}{SRAM}{static RAM}
\newacronym{dram}{DRAM}{dynamic RAM}
\newacronym{hls}{HLS}{high-level synthesis}
\newacronym{pcie}{PCIe}{peripheral component interconnect express}
\newacronym{gops}{GOPS}{giga operations per second}
\newacronym{rnn}{RNN}{recurrent neural network}
\newacronym{lstm}{LSTM}{long short term memory}

%% file: body.tex
{\section{Introduction}
	\label{sec:introduction}}
\glspl{cnn} are used on a daily basis for image \cite{ImageNetClassificationWithDeepConvolutionalNeuralNetworks2012:Krizhevsky}, speech \cite{DNNsForAcousticModelingInSpeechRecognition2012:Hinton} and text recognition \cite{ObjectRecognitionWithGradientBasedLearning1998:Lecunn} and their use and application to different tasks is increasing at a very rapid rate. However, \glspl{cnn} require huge memory storage and bandwidth for weight data and large amounts of computation that would push to extremes the battery, computation and memory in mobile embedded systems. Researchers, \cite{DeepCompression2015:Han,eie2016:Han}, have proposed methods of quantizing and dictionary compressing the weight data to reduce the memory bottleneck and bus bandwidth. Others, \cite{GoingDeeperWithConv2015:szegedy,OptimizingFpgaAccelForCNN2015:Zhang}, have proposed various different \gls{cnn} hardware accelerators implemented in both \glspl{fpga} and \glspl{asic} that may contain hundreds to thousands of parallel hardware \gls{mac} units to increase the computational performance. This increase in computational performance comes at the great expense of power as the \gls{mac} units contain a multiplier, each of which consumes large numbers of logic gates and high power consumption in an \gls{asic} \cite{StudyOfPerformanceComparisonOfDigitalMultipliers2015:Sabeetha}.

\glspl{cnn} are extensively used in an inference mode \cite{ImageNetClassificationWithDeepConvolutionalNeuralNetworks2012:Krizhevsky,VeryDeepConvolutionalNetworksForLargeScaleImageRecognition2014:Simonyan} to infer that, for example, a dog can be found within an image. However the \gls{cnn} must first be trained. Training the \gls{cnn} involves incrementally modifying the ``weight'' values associated with connections in the neural network and retraining until a satisfactory error rate has been achieved \cite{BackpropagationAppliedToHandwrittenZipCodeRecognition1989:LeCun}. At this point the network is considered trained, meaning that no further updates of weight values are required and the trained network can be deployed for inference to the field. In their research Han \MakeLowercase{\textit{et al.}} \citeyear{DeepCompression2015:Han,eie2016:Han} found that in a fully trained \gls{cnn}, similar weight values  occur many times. They proposed scalar quantization of the weight data by clustering around centroids, in order to dictionary compress the weights into bins. They found that between tens to hundreds of weight values were sufficient in network inference whist maintaining the high accuracy rate. They encode the compressed weights with an index that specifies which of the shared weights should be used. This dictionary compression of the weight data reduces the required size and memory bandwidth required for the network. They demonstrate that their weight-shared values can be stored on-chip consuming 5pJ per access rather than in off-chip \gls{dram} which consumed 640pJ per access when implemented on a \gls{cpu} / \gls{gpu} system. Weight sharing does not reduce the number of \gls{mac} operations required; it reduces only the weight data storage and bandwidth requirement.

Building on Han \MakeLowercase{\textit{et al's.}} \citeyear{DeepCompression2015:Han,eie2016:Han} research, we propose a re-architected \gls{mac} circuit of the weight-shared \gls{cnn}  aimed at hardware accelerators. Rather than computing the \gls{sop} in the \gls{mac} directly, we instead count how many times each of the weight indexes appears and store the corresponding image value in a register bin, thus replacing the hardware multipliers with counting, selection and accumulation logic. After this weighted histogram accumulation phase, a post pass multiplication is performed of the accumulated image values in bins with the corresponding weight value of that bin. We call this accelerator optimization the \acrfull{pasm}. To evaluate \gls{pasm} performance we implement \gls{pasm} in a convolution layer of a weight-shared \gls{cnn} accelerator. Where weight bin numbers are small and channel numbers are large, the counting and selection logic can be significantly smaller and lower power than the corresponding multiply-accumulate circuit. We also show that \gls{pasm} is beneficial when implemented in a resource-constrained \gls{fpga} as \gls{pasm} consumes fewer \glspl{bram} and \gls{dsp} units for the \gls{mac} operations in the \gls{fpga}.

The rest of this paper is organized as follows. \hyperref[sec:dnnConv]{Section 2} gives some background on \gls{cnn} accelerators and introduces the \gls{pasm} and how it compares to other \gls{cnn} accelerators. \hyperref[sec:pasmInACnnAccelerator]{Section 3} shows how our \gls{pasm} is implemented in a convolution layer accelerator with examples compared to a weight-shared accelerator. \hyperref[sec:DesignImplementationOfTheWeightSharedWithPASM]{Section 4} describes how a weight-shared-with-\gls{pasm} convolution accelerator is designed and implemented in an \gls{asic} at 45nm clocked at 1GHz and in a Zynq \gls{fpga} clocked at 200MHz. \hyperref[sec:evaluation]{Section 5} presents the experimental results showing latency, power and area projections for both \gls{fpga} and \gls{asic}. \hyperref[sec:relatedWork]{Section 6} reviews related work and \hyperref[sec:conclusion]{Section 7} draws conclusions.


{\section{DNN Convolution with Dictionary-Encoded Weights}
	\label{sec:dnnConv}}
{\subsection{CNN Accelerators}
	\label{subsec:cnnAccelerators}}
A \gls{dnn} contains convolution layers, activation function layers (such as a sigmoid or \gls{relu}) and pooling layers. Up to 90\% of the computation time of a \gls{cnn} is taken up by the convolution layers \cite{hwAccelCnn2010:Farabet}. Within the convolution layer, there are many thousands of \gls{mac} operations, as shown in the pseudo code in Figure \ref{fig:mcmkConvCode}. The convolution operator has an input image of dimensions $IH \times IW$ and $C$ channels and is convolved with $M$ kernels (typically 3 to 832 \cite{GoingDeeperWithConv2015:szegedy}) of dimension $KY \times KX$ and $C$ channels at a stride of $S$ to create an output feature map of $OH \times OW$ and $M$ channels. The loops can be unrolled into parallel \gls{mac} units and implemented in hardware \cite{OptimizingFpgaAccelForCNN2015:Zhang} to accelerate the convolution.

\begin{figure}[t]
	\begin{lstlisting}
image[C][IH][IW], weight[M][C][KY][KX];
outFeat[M][OH][OW];

for (ihIdx=(KY/2); ihIdx<(IH-(KY/2)); ihIdx+=Stride) {
  for (iwIdx=(KX/2); iwIdx<(IW-(KX/2)); iwIdx+=Stride) {
    for (mIdx=0; mIdx<M; mIdx++) {
      summands = 0;
      for (cIdx=0; cIdx<C; cIdx++) {
        for (kyIdx=0; kyIdx<KY; kyIdx++) {
          for (kxIdx=0; kxIdx<KX; kxIdx++) {  
            imVal = image[cIdx][((ihIdx+kyIdx)-(KY/2))][((iwIdx+kxIdx)-(KX/2))];
            kernVal = kernel[mIdx][cIdx][kyIdx][kxIdx];
            summands += imVal * kernVal;
          }
        }
      }
      outFeat[mIdx][ihIdx/Stride][iwIdx/Stride] = summands;
    }
  }
}
	\end{lstlisting}
	\caption{Simplified Pseudo-code of a convolution layer}
	\label{fig:mcmkConvCode}
\end{figure}

\begin{figure}
	\centering
	\includegraphics[width=0.65\linewidth]{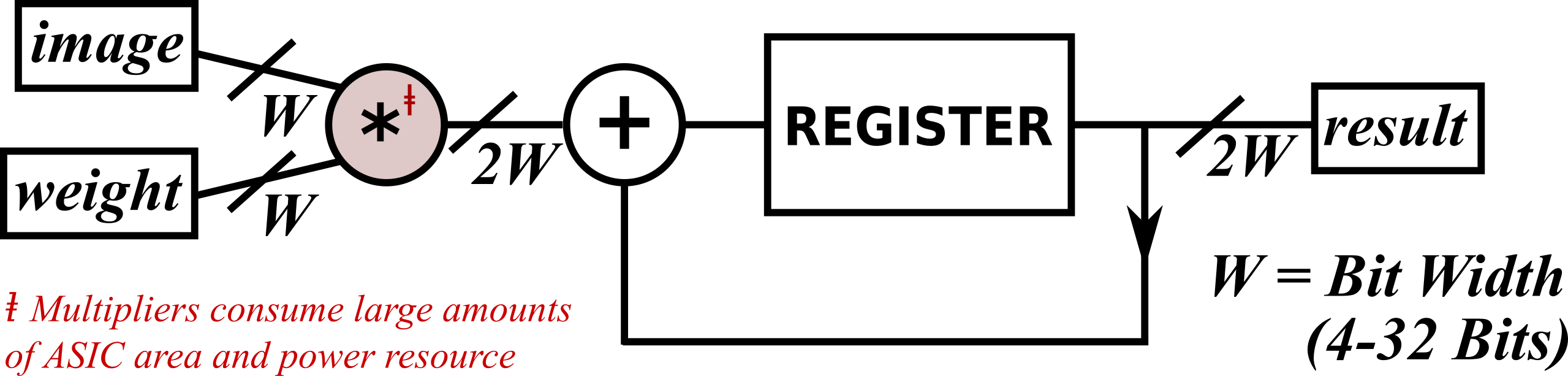}
	\caption{Simple \gls{mac} Block Diagram}
	\label{fig:simpleMacBlockDiagram}
	\vspace{-0.3cm}
\end{figure}

A \gls{mac} unit (see Figure \ref{fig:simpleMacBlockDiagram}) is a sequential circuit that accepts a pair of numeric values (image and weight values) of a predefined bit width and type (e.g. 32-bit fixed point integers), computes their product and accumulates the result in the local accumulator register each clock cycle. The locality of the accumulator register reduces routing complexity and clock delays within the \gls{mac}.

\begin{figure}[t]
	\centering
	\includegraphics[width=0.65\linewidth]{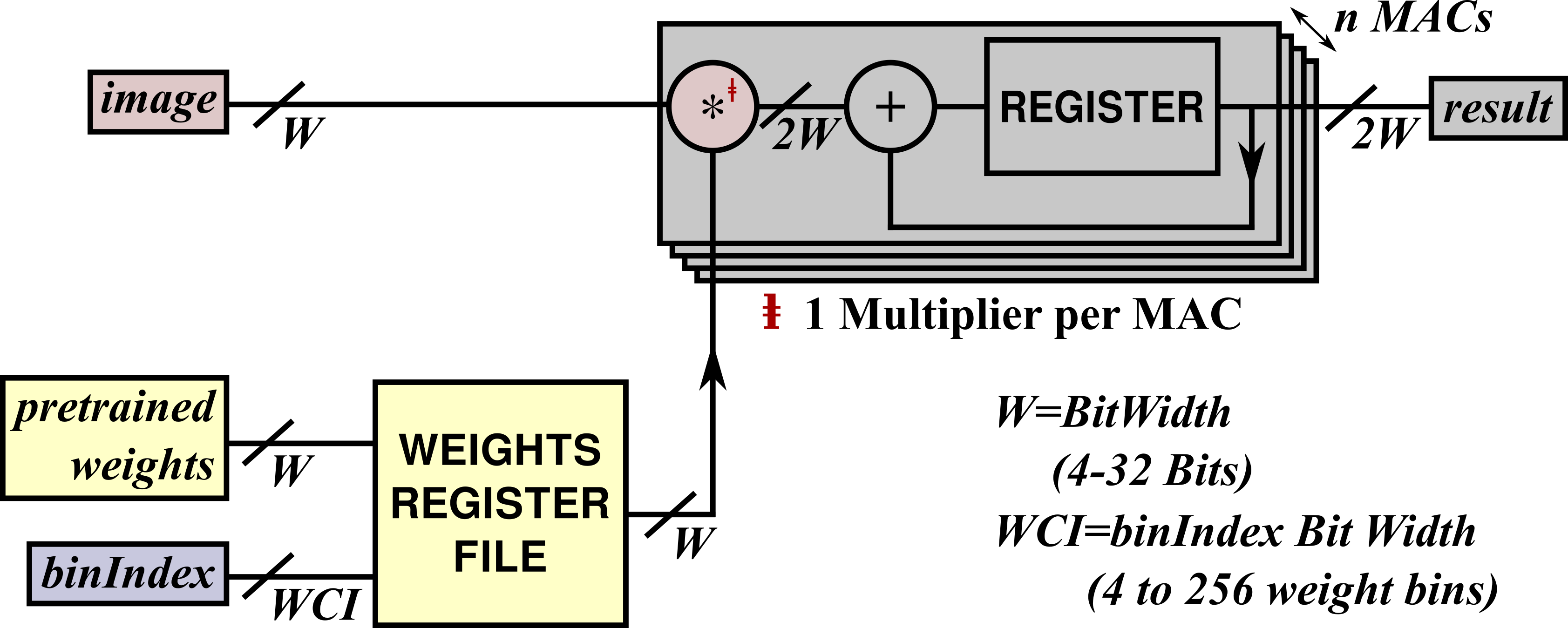}
	\caption{Simplified Weight Shared \gls{mac} Block Diagram.}
	\label{fig:weightSharedRegFileAndMaccBlockDiagram}
\end{figure}

Han \MakeLowercase{\textit{et al.}} \citeyear{DeepCompression2015:Han,eie2016:Han} propose a weight-sharing architecture to reduce the power and memory bandwidth consumption of \glspl{cnn}. They found that similar weight values occur multiple times in a trained \gls{cnn}. By binning the weights and retraining the network with the binned values, they found that just 16 weights were sufficient in many cases. They encode the weights by replacing the original numeric values with a four-bit index that specifies which of the 16 shared weights should be used. This greatly reduces the size of the weight matrices. Figure \ref{fig:weightSharedRegFileAndMaccBlockDiagram} shows simplified weight-sharing decode logic coupled with multiple \gls{mac} units of the \gls{cnn}. When the kernel input is encoded using weight sharing, an extra level of indirection is required to index and access the actual weight value from the weights register file.

\begin{figure}[t]
	\centering
	\includegraphics[width=0.40\linewidth]{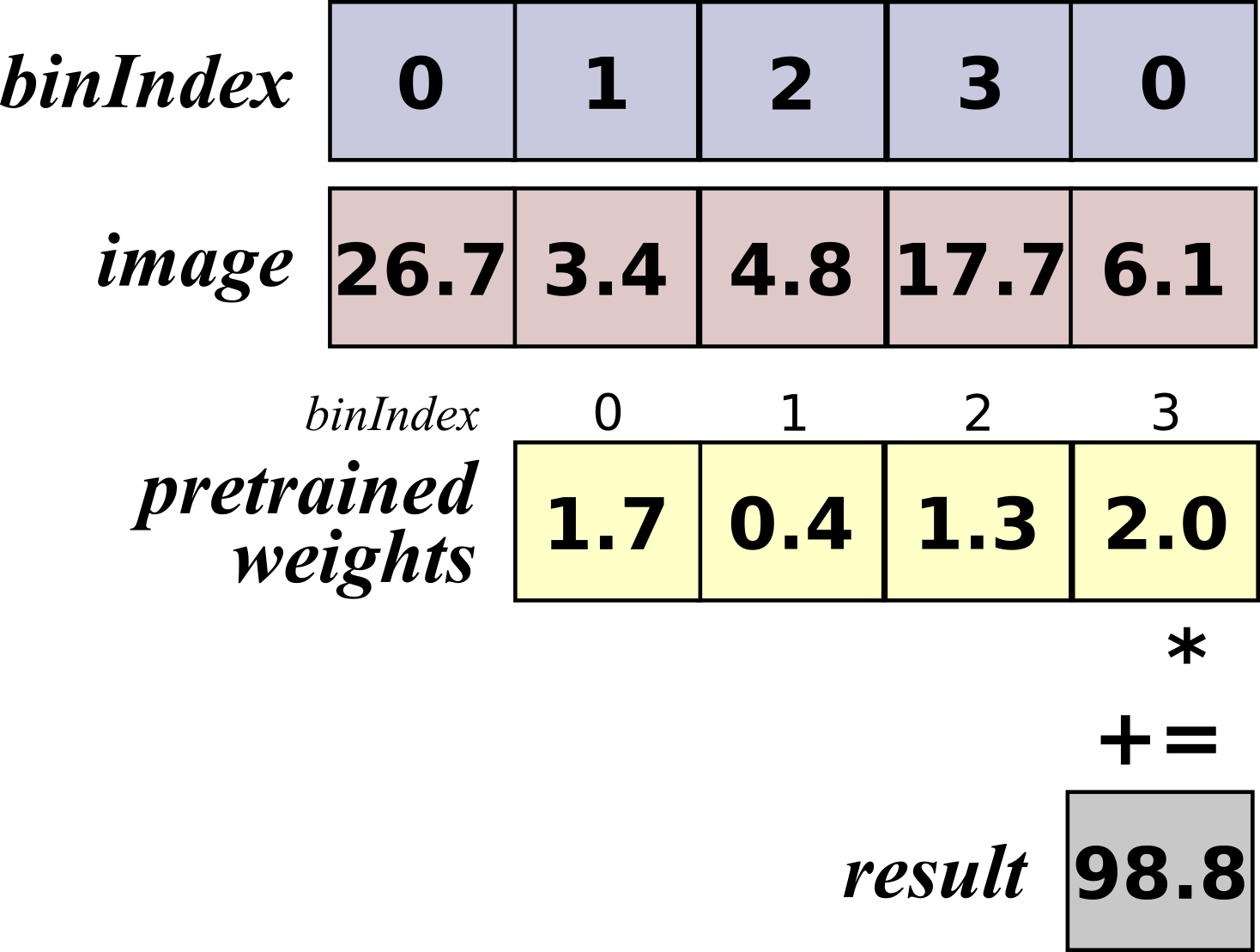}
	\caption{Simplified Weight Shared \gls{mac} example.}
	\label{fig:weightSharedMac}
\end{figure}

Figure \ref{fig:weightSharedMac} shows an example of the weight-shared \gls{mac} in operation. Each \textbf{\textit{image}} value is streamed in, and its corresponding \textbf{\textit{binIndex}} is used to access the pretrained weight against which to multiply and accumulate into the result register. Figure \ref{fig:weightSharedMac} shows how \textbf{\textit{image}} value $26.7$ is multiply-accumulated with the pretrained weight $1.7$ indexed by \textbf{\textit{binIndex}} $0$. Next $3.4$ is multiply-accumulated with the pretrained weight $0.4$ indexed by \textbf{\textit{binIndex}} $1$. This continues until finally multiply-accumulating image value $6.1$ with pretrained weight value $1.7$ of bin $0$ to give:
\[\textbf{\textit{result}}=(26.7 \times 1.7)+(3.4 \times 0.4)+(4.8 \times 1.3)+(17.7 \times 2.0)+(6.1 \times 1.7)=98.8\]

In both a simple \gls{mac} (Figure \ref{fig:simpleMacBlockDiagram}) and a weight-shared \gls{mac} (Figure \ref{fig:weightSharedRegFileAndMaccBlockDiagram}) the multiplier is the most expensive unit in terms of floor area (i.e. large numbers of gates) and power consumption in an \gls{asic} or numbers of \gls{dsp} units in an \gls{fpga}. As a large number of \gls{mac} units are used in a parallel weight-shared \gls{cnn} hardware accelerator, the overall area and power is likely to be large.

Weight sharing is an important factor in implementing \gls{cnn} accelerators in an off-line embedded, low power device. Han \MakeLowercase{\textit{et al.}} \citeyear{DeepCompression2015:Han,eie2016:Han} show that when pruning, quantization, weight-sharing and Huffman coding are all used together in an AlexNet \cite{ImageNetClassificationWithDeepConvolutionalNeuralNetworks2012:Krizhevsky} CNN accelerator, the weight data required is reduced from 240MB to 6.9MB, a compression factor of $35 \times$. Unfortunately, they do not provide results for the effect of weight sharing alone, without these other optimizations. When they apply similar pruning, quantization, weight-sharing and Huffman to the VGG-16 \cite{VeryDeepConvolutionalNetworksForLargeScaleImageRecognition2014:Simonyan} \gls{cnn} accelerator, the weight data is reduced from 552MB to 11.3MB, a $49 \times$ compression ratio. The fully connected layers dominate the model size by $90\%$, but Han \MakeLowercase{\textit{et al.}} \citeyear{DeepCompression2015:Han,eie2016:Han} show that these layers compress by up to $96\%$ of weights pruned in VGG-16 \gls{cnn}. These newly weight-shared \glspl{cnn} run $3 \times$ to $4 \times$ faster on a mobile \gls{gpu} whilst using $3 \times$ to $7 \times$ less energy with no loss in classification accuracy. As the number of free parameters being learnt is reduced in a weight-shared \gls{cnn}, the learning efficiency is greatly increased and allows for better generalization of \glspl{cnn} for vision classification.

The trend is towards increasingly large networks, increasing the number of layers such as ResNet \cite{ResNet2016:He} or increasing the convolution types within each layer such as GoogLeNet \cite{GoingDeeperWithConv2015:szegedy}. Weight sharing is one method that is getting increased research focus to reduce the overall weight data storage and transfer so that the networks can be implemented on off-line mobile devices.

\gls{cnn} hardware accelerators typically use 8-, 16-, 24- or 32-bit fixed point arithmetic \cite{Eyeriss2016a:Chen}. A combinatorial $W$-bit multiplier requires $O(W^2)$ logic gates to implement which makes up a large part of the \gls{mac} unit. Note that sub-quadratic multipliers are possible, but are inefficient for practical values of $W$ \cite{FastIntegerMultiplication:Furer}.

\begin{figure}
	\centering
	\includegraphics[width=1\linewidth]{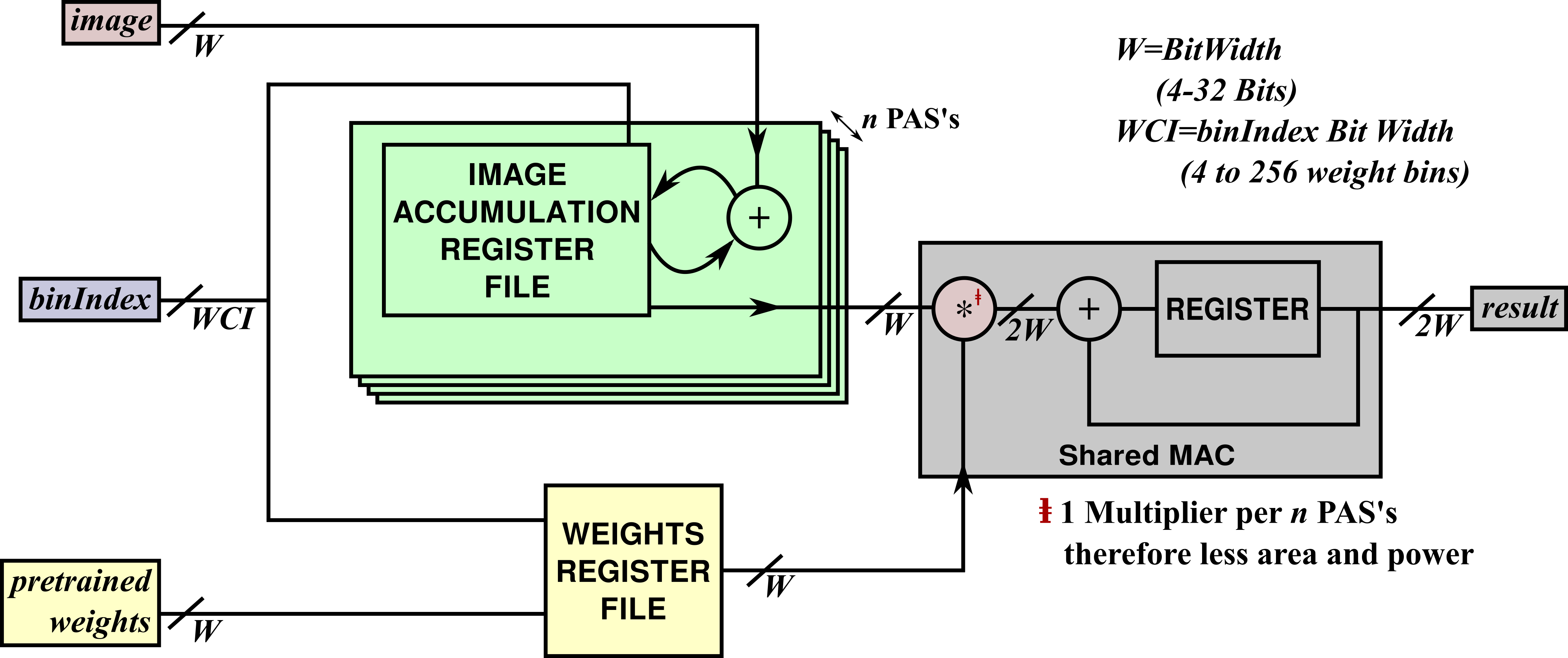}
	\caption{\gls{pasm} showing \gls{pas} unit followed by a shared \gls{mac}.}
	\label{fig:pasmBlockDiagram}
\end{figure}

\vspace{0.5cm}
{\subsection{The PASM Concept}
	\label{subsec:pasm}}
We propose to reduce the area and power consumption of \gls{mac} units by re-architecting the \gls{mac} to do the accumulation first, followed by a shared post-pass multiplication. Our new \gls{pasm} accelerator is shown in Figure \ref{fig:pasmBlockDiagram}. Rather than computing the \gls{sop} in the \gls{mac} directly, \gls{pasm} instead counts how many times each $B$ bin weight-shared index appears and accumulates the corresponding $W$ bit \textbf{\textit{image}} value in the corresponding $B$ weight-shared bin register indexed by the \textbf{\textit{binIndex}}. \gls{pasm} has two phases: (1) accumulate the image values into the weight bins (known as the \gls{pas}) and (2) multiply the binned values with the weights (completing the \gls{pasm}).

\begin{figure}
	\centering
	\subcaptionbox{Phase 1: As each image value is streamed in, its associated bin index is also streamed so that the image values can be accumulated into correct bins.\label{fig:pasPhase}}{\includegraphics[width=0.30\linewidth]{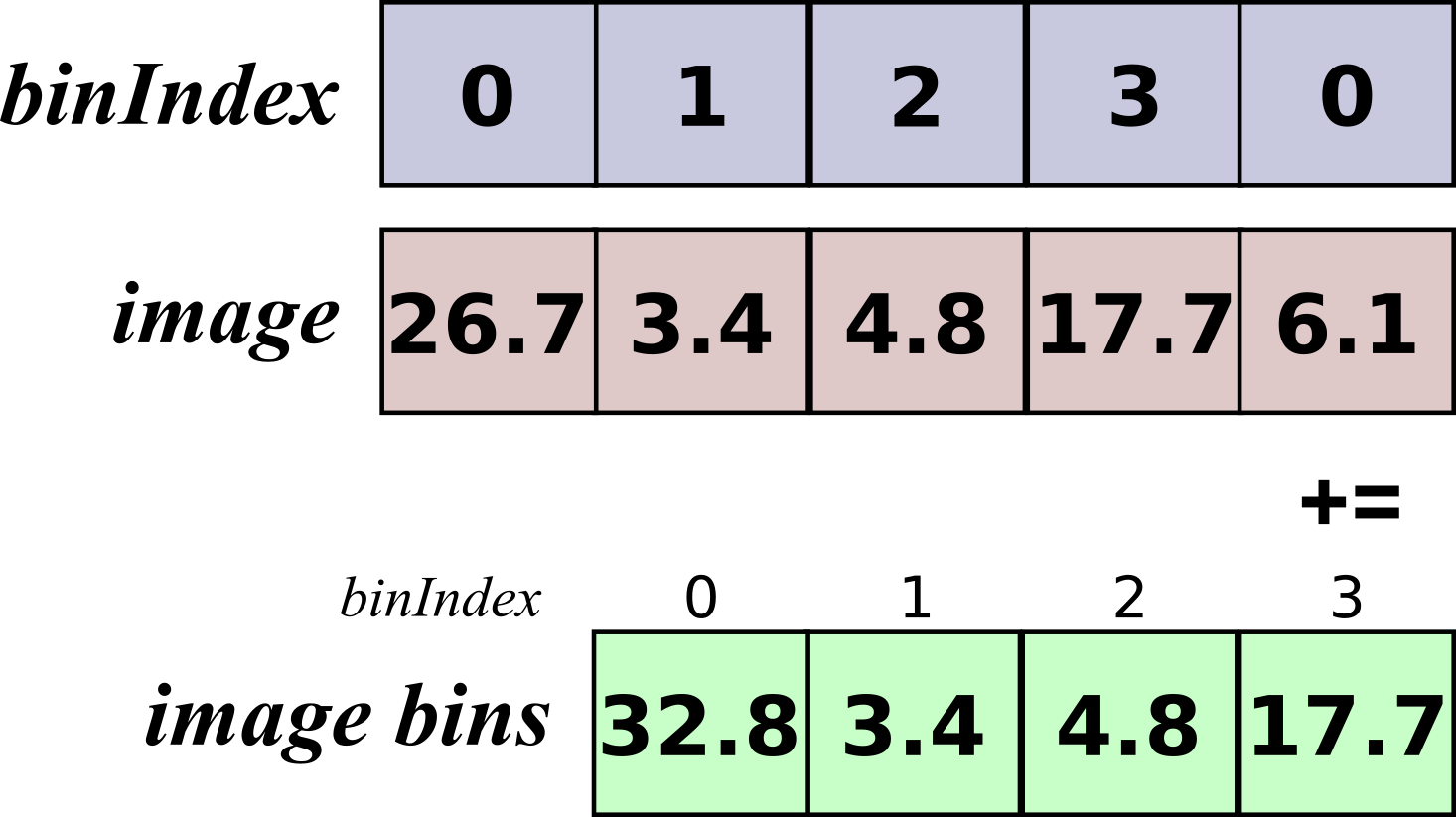}}\hspace{3em}%
	\subcaptionbox{Phase 2: Each bin accumulated value is multiplied with its corresponding pre-trained weight value to produce the final  result.\label{fig:mulPhase}}{\includegraphics[width=0.30\linewidth]{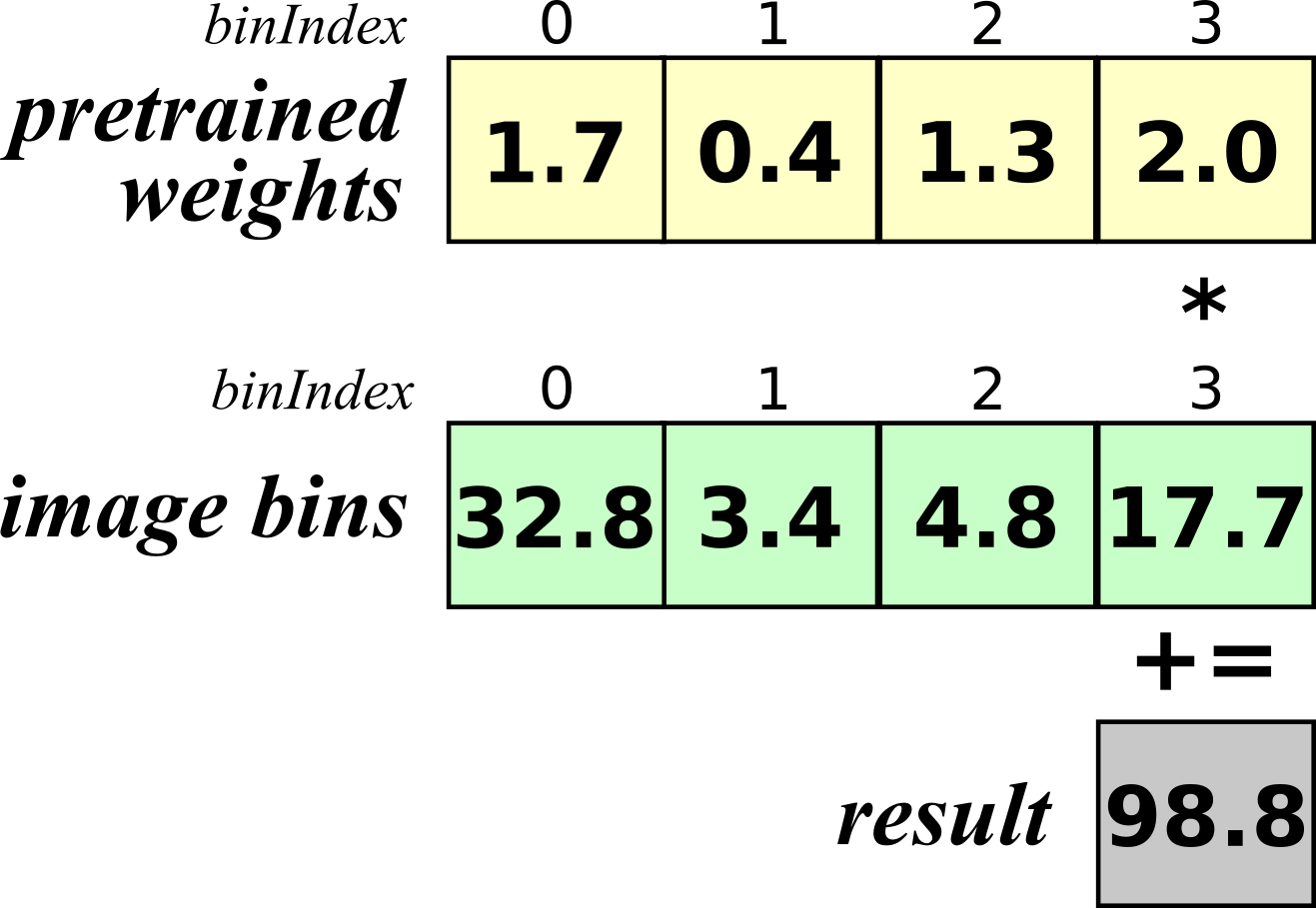}}
	\caption{PASM in Operation}
	\label{fig:pasmInOperation}
\end{figure}

Figure \ref{fig:pasPhase} shows an example of the accumulation phase. Our \gls{pas} unit is a sequential circuit that consumes a pair of inputs each cycle. One input is an \textbf{\textit{image}} value, and the other is the \textbf{\textit{binIndex}} of the weight value in the dictionary of weight encodings. The \gls{pas} unit has a set of $B$ accumulators, one for each entry in the dictionary of weight encodings. The accumulators are initially set to zero. Each time the \gls{pas} consumes an input pair, it adds the \textbf{\textit{image}} value to the accumulator with index \textbf{\textit{binIndex}}. For example, when the leftmost pair of inputs in Figure \ref{fig:pasPhase} are consumed, the \textbf{\textit{image}} value $26.7$ is added onto accumulator numbered \textbf{\textit{binIndex}} $=0$. Next $3.4$ is accumulated into bin $1$. This continues until finally accumulating $6.1$ into bin $0$ to give $26.7+6.1=32.8$. This accumulated result tells us that the weight stored in dictionary location $0$ has been paired with an accumulated \textbf{\textit{image}} value of $32.8$. For the accumulation phase, the actual weight value stored in dictionary location $0$ does not matter. We are simply computing a weighted histogram of the dictionary weight indices.

In the second phase, the histogram of weight indices is combined with the actual weight values to compute the result of the sequence of multiply-accumulate operations. Figure \ref{fig:mulPhase} demonstrates the multiply phase, multiplying-accumulating bin $0$ \textbf{\textit{pretrained weight}} with bin $0$ accumulated \textbf{\textit{image}} value, giving $32.8 \times 1.7=55.76$. The contents of \textbf{\textit{pretrained weight}} bin $1$ is multiplied-accumulated with \textbf{\textit{image}} bin $1$ value and so on until all the corresponding bins are multiplied-accumulated into the \textbf{\textit{result}} register, giving $98.8$, the same result found by the weight shared \gls{mac}, Figure \ref{fig:weightSharedMac}.

This second, multiply stage can be implemented using a traditional \gls{mac} unit that is shared between several \gls{pas} units. Several \gls{mac} units can be replaced by the same number of \gls{pas} units sharing a single \gls{mac}. For example, consider the case where we must compute many multiply-accumulate sequences, where each sequence consumes $1024$ pairs (image and weight) of values. A fully-pipelined \gls{mac} unit is a sequential circuit that will typically require a little over $1024$ cycles to compute the result.

If we have four such \gls{mac} units, we can compute four such results in parallel, again in around $1024$ cycles. If the weight data has been quantized and dictionary encoded to just, say, $16$ values, then we could use \gls{pas} units with $B=16$ bins to perform the accumulate phase of the \gls{pasm} computation. Four such fully-pipelined \gls{pas} units could perform the accumulation phase in around $1024$ cycles. However, the accumulation phase of the \gls{pasm} does not give us the complete answer. We also need to perform the multiply phase, which involves multiplying and accumulating $B=16$ values in this example. If each \gls{pas} unit had its own \gls{mac} unit, then the multiply phase would take around $16$ cycles for a total of $1024+16=1040$ cycles for the entire multiply accumulate operation. However, in this example, the four parallel \gls{pas} units share a single \gls{mac} unit with the result that the total time will be $1024 + 4 \times 16 = 1088$ cycles. \gls{pasm} can have higher throughput when compared to the standard \gls{mac} due to the \gls{pas} units being much smaller than the \gls{mac} for small values of $B$, up to about $B=16$.

\vspace{0.4cm}
{\subsection{PASM accelerator}
	\label{subsec:pasmAccelerator}}

Table \ref{table:complexity} compares the gate counts of the sub components of a simple \gls{mac}, a weight-shared \gls{mac} and a \gls{pas}. The \textit{gates} column shows the circuit complexity in gates of each sub-component, assuming fixed-point arithmetic.  The bit-width of the data is $W$ and the number of bins is $B$ in the weight-shared designs. For example, a simple \gls{mac} unit contains an adder ($O(W)$ gates), a multiplier ($O(W)^2$) gates) and a register ($O(W)$ gates). A weight shared \gls{mac} also needs a small register file with $B$ entries to allow fast mapping of encoded weight indices to shared weights. The \gls{pas} needs a read and write port due to the interim storage of the accumulation results that need to be read by the post pass multiplier whereas the \gls{mac} only needs a write port.

From Table \ref{table:complexity} we can also see that the efficiency of \gls{pas} depends on a weight-sharing scheme where the number of bins, $B$, is much less than the total number of possible values that can be represented by a weight value, that is $2^W$. For example, if we consider the case of $W=16$, then in the absence of weight sharing, a \gls{pas} would need to deal with the possibility of $2^{16}$ different weight values, requiring $2^{16}$ separate bins.  The hardware area of these bins is likely to be prohibitive. Therefore, \gls{pas} is effective where the number of bins is much lower than $2^W$.

\begin{table}[H]
	\centering
	\caption{Complexity of \gls{mac}, Weight-shared \gls{mac} and \gls{pas}}
	\begin{tabular}{l|c|c|c|c}
		\hline
		Sub Component   		& Gates            & Simple & Weight Shared     & PAS \\
		&                  & MAC    & MAC               &     \\
		\hline
		Adder			        & $O(W)$           & 1      & 1      			& 1   \\
		Multiplier          	& $O(W^2)$         & 1      & 1      			&     \\
		Weight Register			& $O(W)$           & 0      & $B$      			&	  \\
		Accumulation Register	& $O(W)$           & 1	    & 1					& $B$ \\
		File Port				& $O(WB)$          &        & 1      	        & 2   \\
		\hline
	\end{tabular}
	\label{table:complexity}
\end{table}

{\subsection{Evaluation of PASM as a stand-alone unit}
	\label{subsec:evalPasmStandAlone}}
We design an accelerator unit to perform a simplified version of the accumulations in Figure \ref{fig:pasmBlockDiagram}. Our accelerator accepts 4 image inputs and 4 shared-weight inputs each cycle and uses them to compute 16 separate \gls{mac} operations each cycle. The weight-shared version performs these operations on 16 weight-shared \gls{mac} units (16-\gls{mac}). Our proposed \gls{pasm} unit has 16 \gls{pas} units and uses 4 \gls{mac} units for post-pass multiplication (16-\gls{pas}-4-\gls{mac}). Both the weight-shared and weight-shared-with-\gls{pasm} accelerators are coded in Verilog 2001 and synthesized to a flat netlist at 100MHz with a short 0.1ns clock transition time targeted at a 45nm process \gls{asic}. We measure and compare the timing, power and gate count in both designs for the same corresponding bit widths and same numbers of weight bins.

The standard 16-\gls{mac} and the proposed 16-\gls{pas}-4-\gls{mac} each have $W$ bit \textbf{\textit{image}} and \textbf{\textit{weight}} inputs and the 16-\gls{pas}-4-\gls{mac} has a $WCI$ bit \textbf{\textit{binIndex}} input to index into the $B=2^{wci}$ weight bins. The designs are coded using integer/fixed point precision numbers. Both versions are synthesized to produced a gate level netlist and timing constraints designed using \acrfull{sdc} \cite{FpgaXdcTiming2015:Gangadharan} so that both designs meet timing at 100MHz.

Cadence Genus (version 15.20 - 15.20-p004\_1) is used for synthesizing the \gls{rtl} into the OSU FreePDK 45nm process \gls{asic} and applying the constraints in order to meet timing. Genus supplies commands for reporting approximate timing, gate count and power consumption of the designs at the post-synthesis stage. The ``report timing'', ``report gates'' and ``report power'' commands of Cadence Genus are used to obtain the results for both 16-\gls{mac} and 16-\gls{pas}-4-\gls{mac} accelerators. Graphs of the gate count and power consumption results are produced for the two different designs at different bit widths and different numbers of weight bins, showing that the \gls{pasm} is consistently smaller and more efficient than the weight-sharing \gls{mac}.

\begin{figure}
	\centering
	\includegraphics[width=1\linewidth]{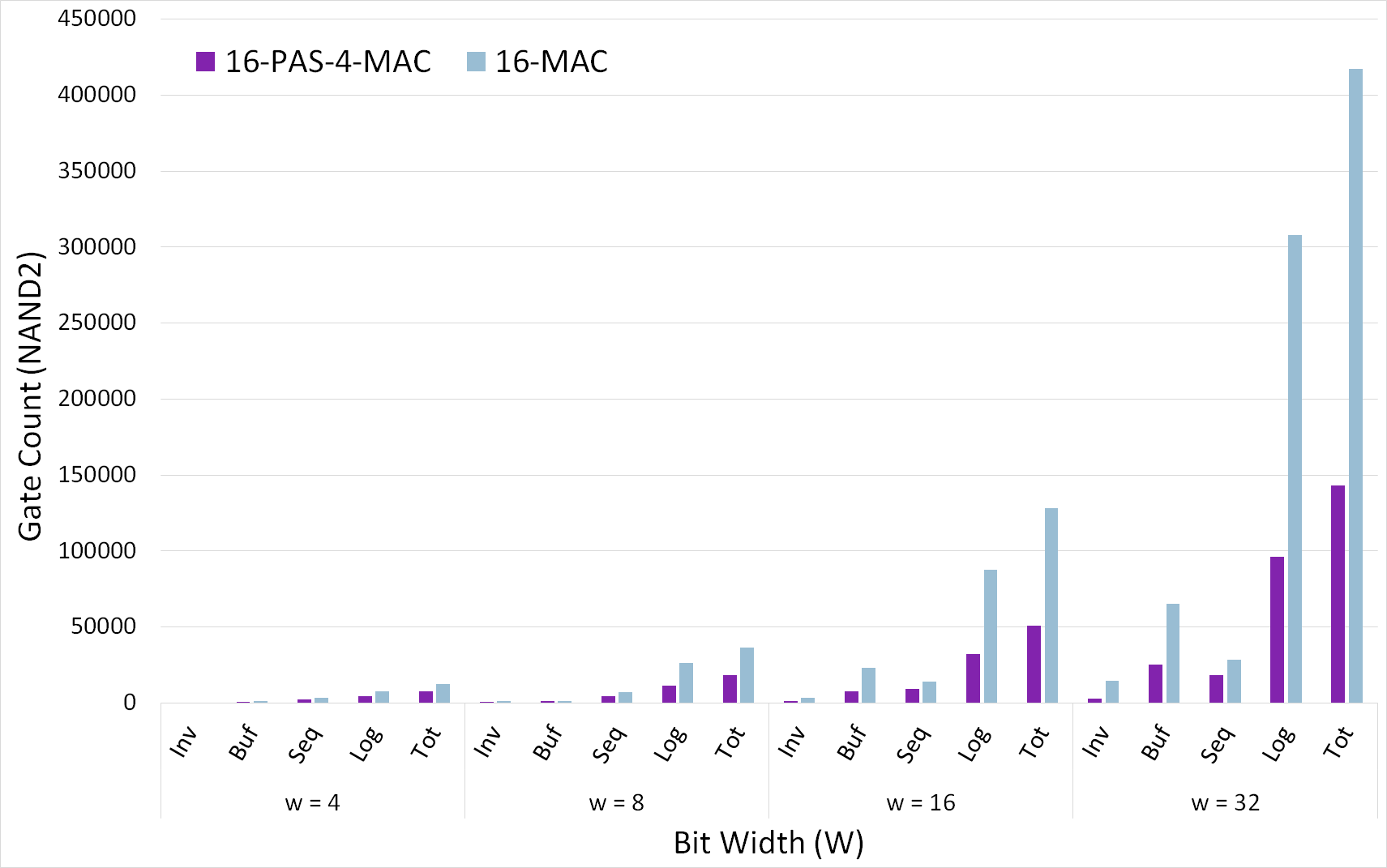}
	\caption{Logic gate count comparisons (in NAND2X1 gates) for $W=4, 8, 16, 32$-bits wide 16-\gls{mac} and 16-\gls{pas}-4-\gls{mac} for $B=16$ weight bins - \textbf{lower is better}}
	\label{fig:asicUtilizationComparisonWidth}
\end{figure}

Figure \ref{fig:asicUtilizationComparisonWidth} shows comparisons of the logic resource requirements of a $B=16$ shared-weight-bin 16-\gls{pas}-4-\gls{mac} and 16-\gls{mac} for varying $W$ bit widths. Gate counts are normalized to a NAND2X1 gate. The \gls{pasm} uses significantly fewer logic gates. For example, for $W=32$ bits wide the 16-\gls{pas}-4-\gls{mac} is 35\% smaller in sequential logic, 78\% smaller in inverters, 61\% smaller in buffers and 68\% smaller in logic, an overall 66\% saving in total logic gates. The \gls{pasm} requires more accumulators for the $B$-entry register file, but otherwise overall resource requirements are significantly lower than that of the \gls{mac}.

\begin{figure}
	\centering
	\includegraphics[width=1\linewidth]{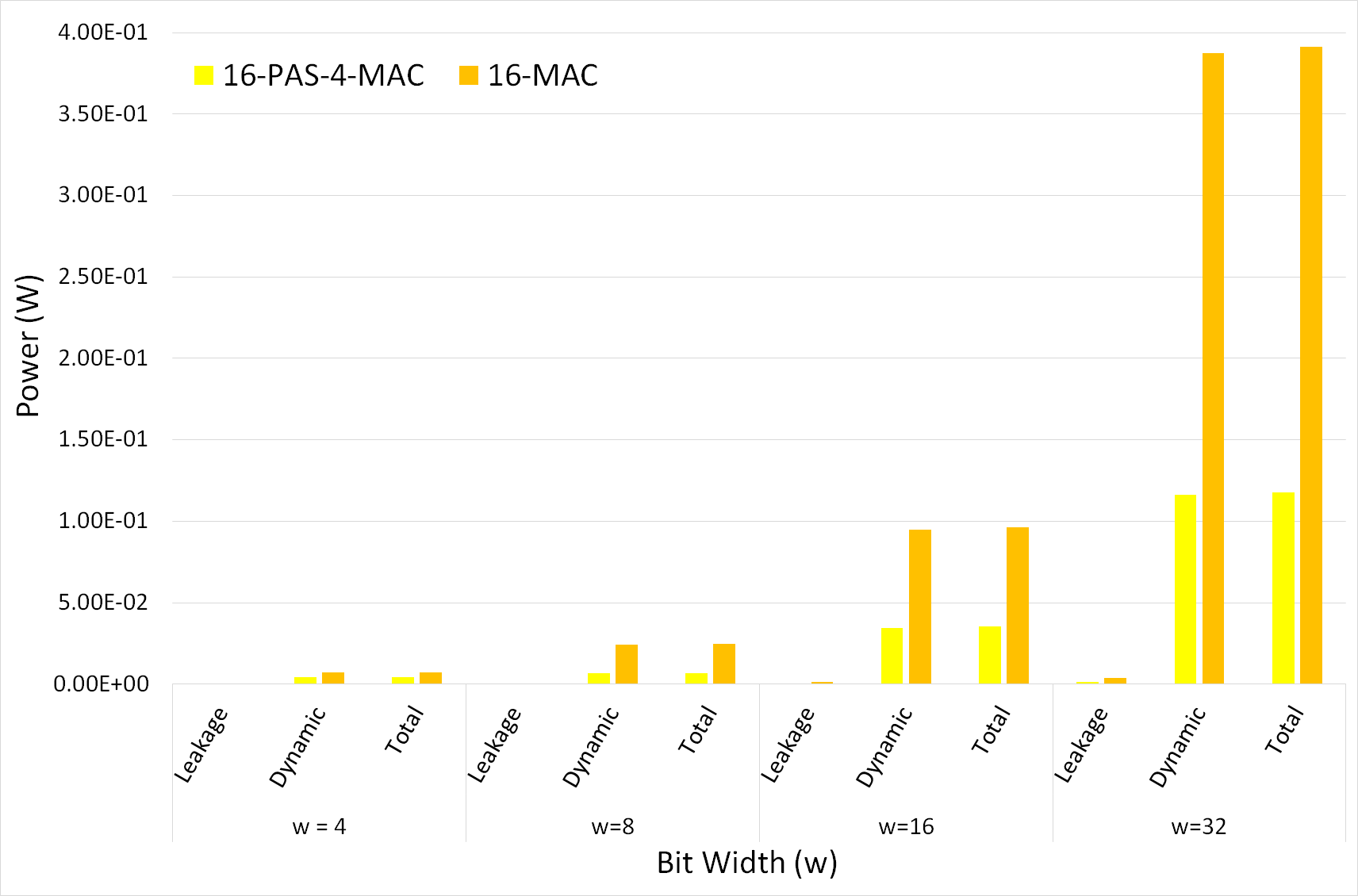}
	\caption{Power consumption (in W) comparisons for $W=4, 8, 16, 32$-bits wide 16-\gls{mac} and 16-\gls{pas}-4-\gls{mac} for $B=16$ weight bins - \textbf{lower is better}}
	\label{fig:asicTotalDynamicPowerComparisonWidth}
\end{figure}

Figure \ref{fig:asicTotalDynamicPowerComparisonWidth} shows comparisons of power consumption of the accelerators. 16-\gls{pas}-4-\gls{mac}'s power is lower than the weight-shared 16-\glspl{mac} and the gap grows with increasing $W$-bit width. For example, for the $W=32$-bit versions of each design, the 16-\gls{pas}-4-\gls{mac} consumes 60\% less leakage power, 70\% less dynamic power and 70\% less total power than that of the 16-\gls{mac} version.

\begin{figure}
	\centering
	\includegraphics[width=1\linewidth]{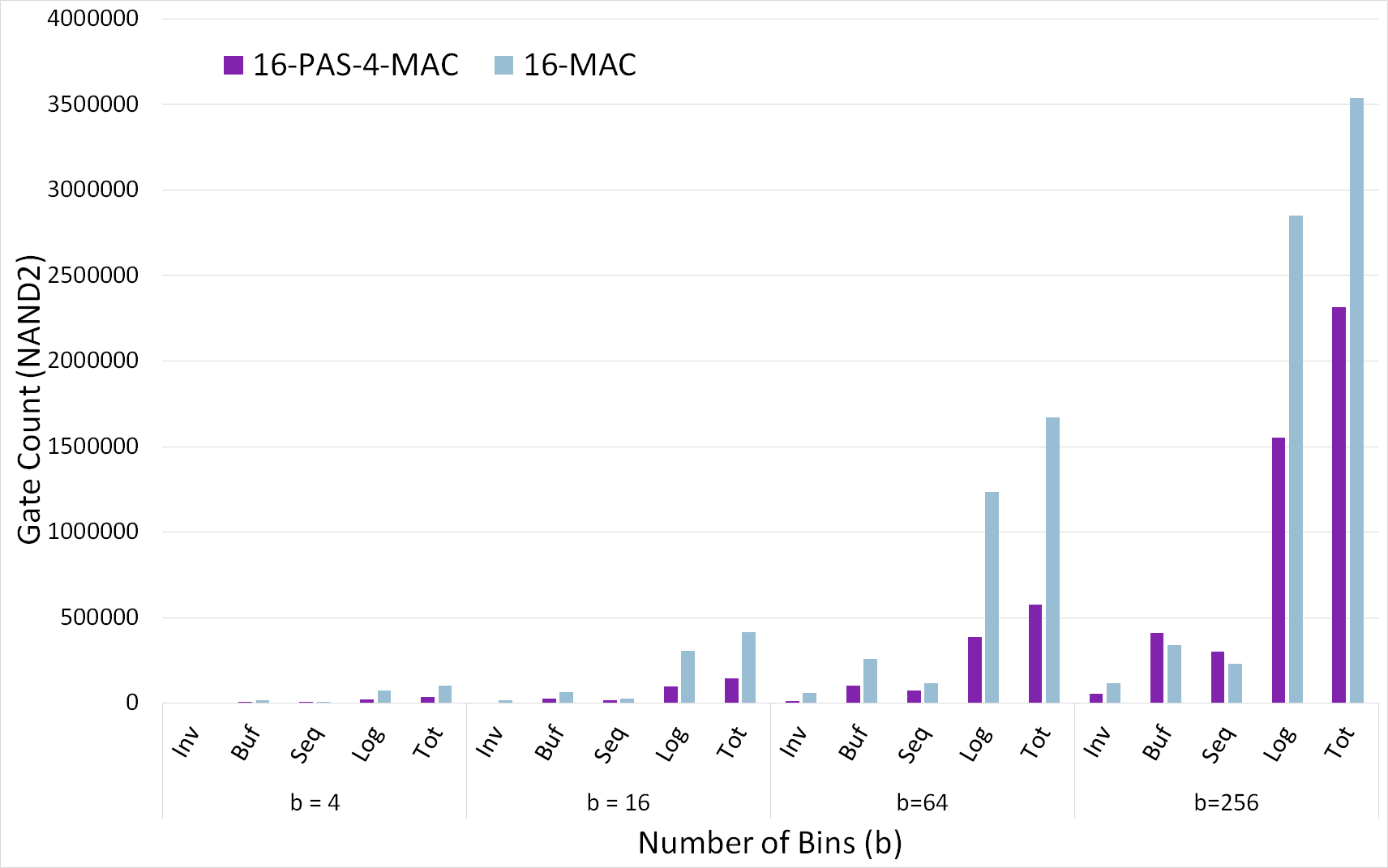}
	\caption{Logic gate counts comparisons  (in NAND2X1 gates) for $B=4, 16, 64, 256$ weight bins for a 16-\gls{mac} and 16-\gls{pas}-4-\gls{mac} for $W=32$-bit width - \textbf{lower is better}.}
	\label{fig:asicUtilizationComparisonDepth}
\end{figure}

Figure \ref{fig:asicUtilizationComparisonDepth} shows the effect of varying the number of bins from $B=4$ to $B=256$, with gate counts normalized to a NAND2X1. For bit width $W=32$ and $B=16$ bins the 16-\gls{pas}-4-\gls{mac} utilization has 35\% fewer sequential gates, 78\% fewer inverters, 62\% fewer buffers and 69\% fewer logic and 66\% less total logic gates compared to the 16-\gls{mac} design. However, at $B=256$, \gls{pasm} registers and buffers are less efficient than the \gls{mac}.

\begin{figure}
	\centering
	\includegraphics[width=1\linewidth]{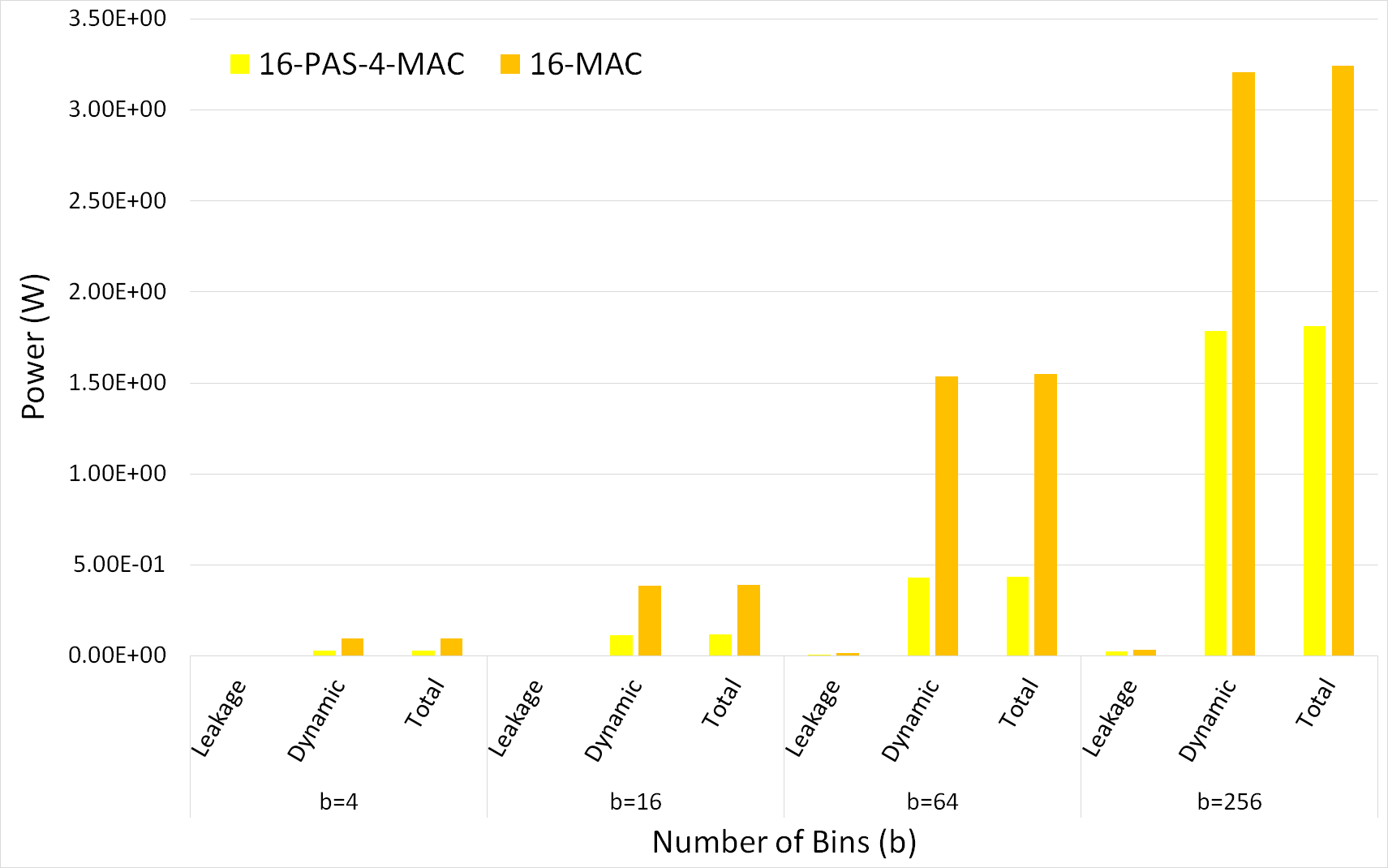}
	\caption{Power consumption (in W) comparisons for $B=4, 16, 64, 256$ weight bins deep 16-\gls{mac} and 16-\gls{pas}-4-\gls{mac} for $W=32$-bit width - \textbf{lower is better}.}
	\label{fig:asicTotalDynamicPowerComparisonDepth}
\end{figure}

The 16-\gls{pas}-4-\gls{mac} also consumes 61\% less leakage power, 70\% less dynamic power and 70\% less total power (Figure \ref{fig:asicTotalDynamicPowerComparisonDepth}). More details can be found in our original paper, \cite{LowComplexityMacForWeightSharingCnn2017:Garland}.

\vspace{0.4cm}
\section{PASM in a CNN Accelerator}
\label{sec:pasmInACnnAccelerator}
\begin{figure}
	\centering
	\includegraphics[width=0.95\linewidth]{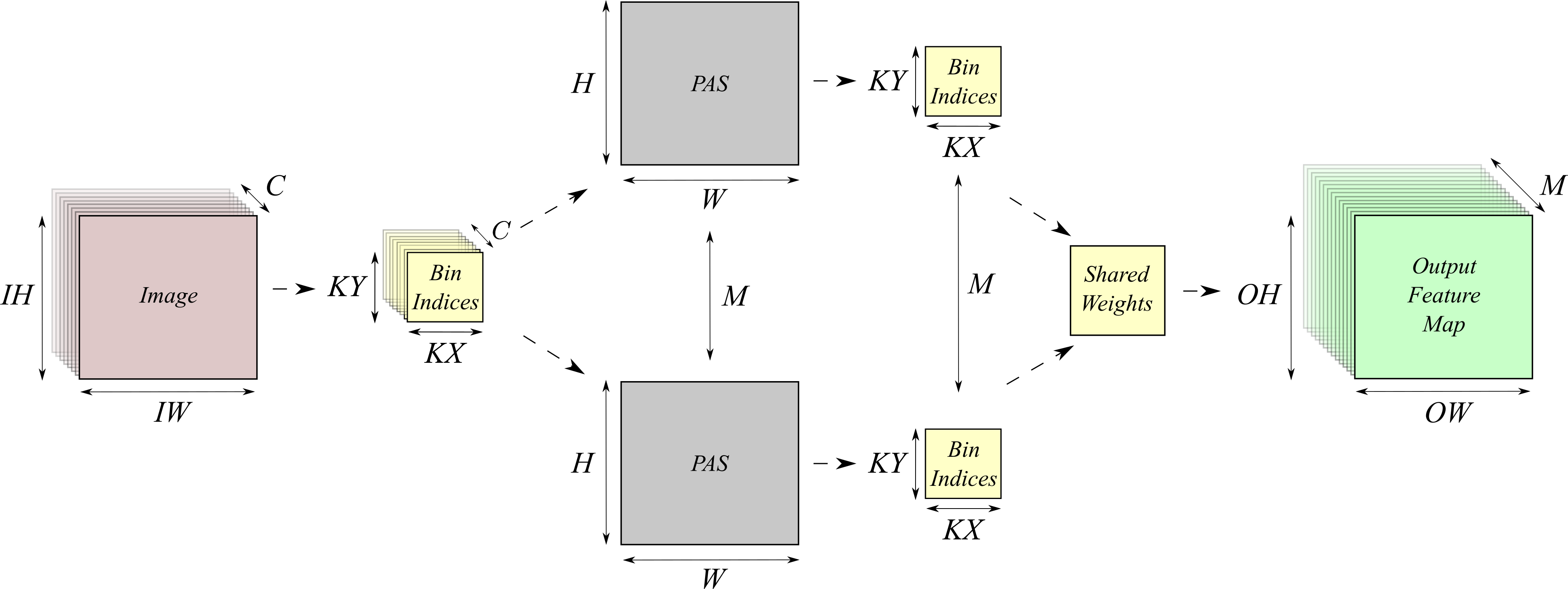}
	\caption{Example of a Simplified Weight Shared Convolution with \gls{pasm}.}
	\label{fig:weightSharedExampleWithPasm}
\end{figure}

In this paper, we asked the question would \gls{pasm} offer similar power and area savings when implemented in a layer of a \gls{cnn} accelerator and how would it affect performance of the convolution accelerator? We attempt to answer this by implementing \gls{pasm} in a weight-shared convolution layer accelerator and evaluate and compare its latency, power and area performance with a weight-shared convolution accelerator and baseline both against a non-weight shared convolution accelerator for the same clock speed. Figure \ref{fig:weightSharedExampleWithPasm} shows how, when \gls{pasm} is implemented in a weight-shared convolution accelerator, multiple \gls{pas} units are created in parallel to accelerate the accumulation of $C \times IH \times IW$ \textbf{\textit{image}} data into the corresponding $B$ bin registers. Multiplexers are created to expand and parallelize the \textbf{\textit{image}} and \textbf{\textit{binIndex}} data and demultiplexers then combine the \gls{pas} outputs for the post-pass \gls{mac}. The post-pass \gls{mac} multiplies and accumulates the binned \textbf{\textit{image}} data with the corresponding $M \times C \times KX \times KY$ shared-weight value into the $M \times IH \times IW$ \textbf{\textit{outFeat}}.

The \textbf{\textit{image}} data of $C \times IH \times IW$ are buffered in registers, \textbf{\textit{weight}} data of $M \times C \times KX \times KY$ are buffered in shared weight registers, the \textbf{\textit{binIndex}} data up to 16 values are registered and finally the output feature map of $M \times IH \times IW$ is registered in an \textbf{\textit{outFeat}} register file. This allows for greater locality and reuse of the data. 

As can be seen from Table \ref{table:complexity} and Table \ref{table:channelsKernelsMatrix}, the \gls{pasm} is only efficient when the number of \gls{pas} units created is much smaller than the number of items to accumulate, i.e. the \gls{pasm} is efficient only where the number of bins, $B$, is much smaller that the number of pairs of inputs to be multiplied and summed, $C \times K \times K$. In the absence of quantization and weight-sharing, the \gls{pasm} would not be viable. For example, if we tried to use \gls{pasm} for 16-bit weight values without using quantization or weight-sharing, then we would need 2\textsuperscript{16} bins in the \gls{pasm}. A \gls{pasm} unit with so many bins would not be competitive with a conventional MAC unit.

Any weight-shared network such as a weight-shared AlexNet \cite{ImageNetClassificationWithDeepConvolutionalNeuralNetworks2012:Krizhevsky}, weight-shared VGG \cite{VeryDeepConvolutionalNetworksForLargeScaleImageRecognition2014:Simonyan} or weight-shared GoogLeNet \cite{GoingDeeperWithConv2015:szegedy}, and more generally regional \glspl{cnn}, \glspl{rnn} and \glspl{lstm} are possible good candidates for the use of \gls{pasm}, although the evaluation in these networks is beyond the scope of this paper.

\begin{figure}[H]
	\centering
	\includegraphics[width=.8\linewidth]{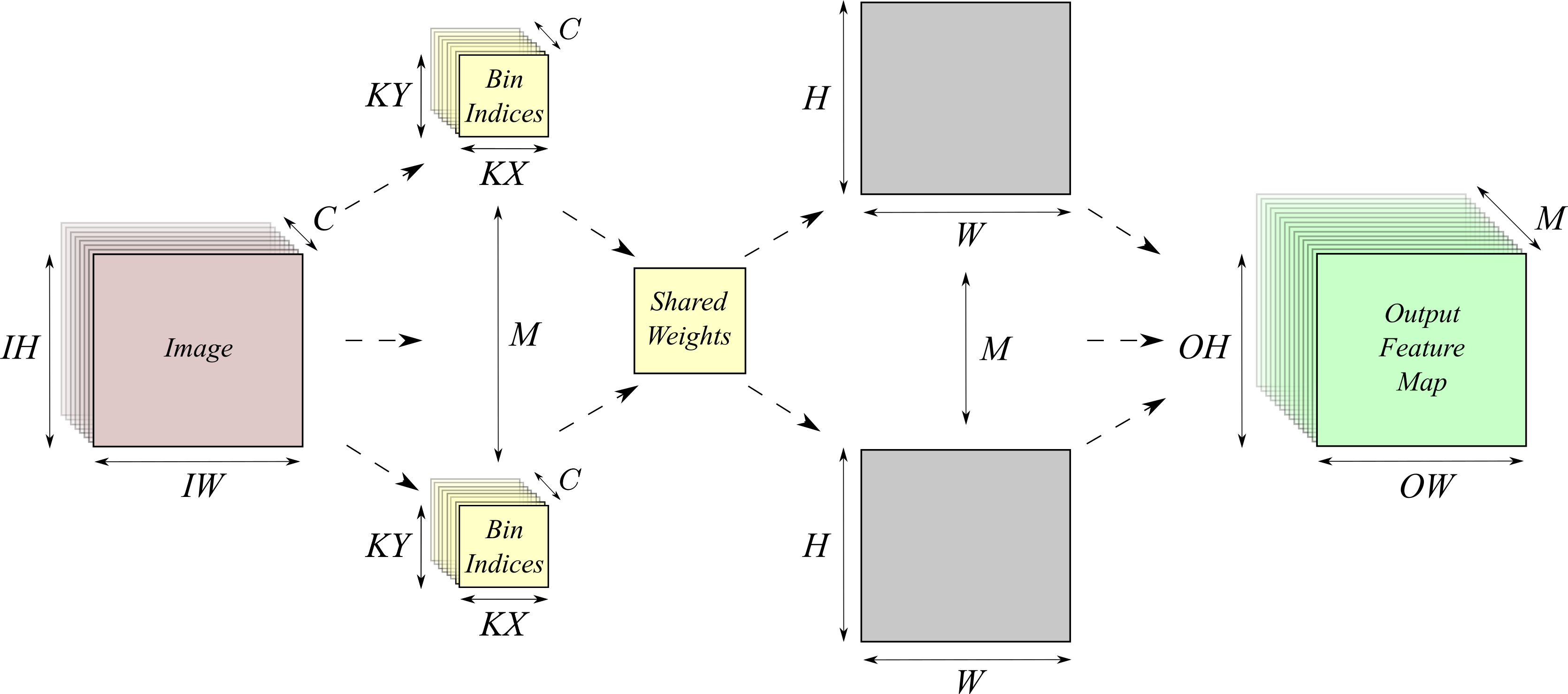}
	\caption{Example of a Simplified Weight-Shared Convolution.}
	\label{fig:weightSharedExample}
\end{figure}

{\subsection{Examples}
	\label{subsec:examples}}
For a simplified weight-shared accelerator, Figure \ref{fig:weightSharedExample}, each kernel channel is `slid' across the corresponding image channel, multiplying and accumulating each of the pixel values with the kernel's pre-trained weight-shared values into the corresponding interim feature map channel. Each of the interim feature map channels is then `stacked' to produce the output feature map.

Now assume a simplified weight-shared-with-\gls{pasm} accelerator with the same number of channels and kernels, Figure \ref{fig:weightSharedExampleWithPasm}. Again each kernel channel is `slid' across the corresponding image channel, however, the `kernel' contains bin indices that address the interim feature map bin into which the image pixel values are accumulated. After all the image channels have been accumulated into the image bins of the interim feature map, the bin indices are `slid' across the interim feature map, multiplying each of the accumulated image values with the corresponding kernel's indexed pre-trained weight-shared values, and accumulated into the associated output feature map channel.

Figure \ref{fig:mcmkPasmConvCode} shows the simplified System C code for weight-shared-with-\gls{pasm} implemented within a convolution layer. It demonstrates an \textit{\textbf{image}} of $C \times IH \times IW$, a \textbf{\textit{kernel}} of $M \times C \times KY \times KX$, with $B$ \textbf{\textit{weight}} bins, a \textbf{\textit{stride}} of $S$ and an \textbf{\textit{outFeat}} of $M \times OH \times OW$.

\begin{figure}
	\begin{lstlisting}
bi[C][KY][KX], sk[B], bias[M], image[C][IH][IW], imageBin[B], outFeat[M][OH][OW];
#pragma HLS ARRAY_PARTITION variable=imageBin complete dim=1
#pragma HLS ALLOCATION instances=mul limit=1 operation
for (ihIdx=(KY/2); ihIdx<(IH-(KY/2)); ihIdx+=Stride) {
  for (iwIdx=(KX/2); iwIdx<(IW-(KX/2)); iwIdx+=Stride) {
    for (mIdx=0; mIdx<M; mIdx++) {
#pragma HLS PIPELINE II=1 rewind
// Reset the imageBin register file
      for (bin=0; bin<B;bin++) {
#pragma HLS UNROLL
#pragma HLS LOOP_MERGE
        imageBin[bin]=0;
      }
      
      binIdx=0;
      // For each channels, stride the kernel sized bin indices over the
      // image and accumulate the image value in the corresponding imageBin PAS
      for (cIdx=0; cIdx<C; cIdx++) {
        for (kyIdx=0; kyIdx<KY; kyIdx++) {
          for (kxIdx=0; kxIdx<KX; kxIdx++) {
            imVal=image[cIdx][((ihIdx+kyIdx)-(KY/2))][((iwIdx+kxIdx)-(KX/2))];
            binIdx=bi[cIdx][kyIdx][kxIdx];
            imageBin[binIdx] += imVal;
          }  // end for (kxIdx=0; ...
        }  // end for (kyIdx=0; ...
      }  // end for (cIdx=0; ...
	
      // Once looped over all the channels, stride the kernel sized bin indices
      // over the PAS and multiply with the corresponding shared-weight value.
      cIdx=0;
      for (kyIdx=0; kyIdx<KY; kyIdx++) {
        for (kxIdx=0; kxIdx<KX; kxIdx++) {
          mul[cIdx][kyIdx][kxIdx] =imageBin[bi[cIdx][kyIdx][kxIdx]] * 
                                                 sk[bi[cIdx][kyIdx][kxIdx]];
        }  // end for (kxIdx=0; ...
      }  // end for (kyIdx=0; ...

      // Sum all the channel's together into each ouptut feature map channel 
      for (cIdx=0; cIdx<C; cIdx++) {
        for (kyIdx=0; kyIdx<KY; kyIdx++) {
          for (kxIdx=0; kxIdx<KX; kxIdx++) {
            outFeat[mIdx][ihIdx/Stride][iwIdx/Stride] += mul[cIdx][kyIdx][kxIdx];
          }  // end for (kxIdx=0; ...
        }  // end for for (kyIdx=0; ...
      }  // end for (cIdx=0; ...
    }  // for (mIdx=0; ...
  }  // for (iwIdx=(KX/2); ...
}  // for (ihIdx=(KY/2); ...
	\end{lstlisting}
	\caption{Simplified System-C Code for the weight-shared-with-\gls{pasm} convolution}
	\label{fig:mcmkPasmConvCode}
\end{figure}


\vspace{0.4cm}
{\section{Design and Implementation of the PASM CNN Accelerator}
	\label{sec:DesignImplementationOfTheWeightSharedWithPASM}}
For comparison, three versions of the accelerator, a non-weight-shared, a weight-shared and a weight-shared-with-\gls{pasm} accelerator are designed and synthesized. The accelerators are coded in SystemC which allows the designs to be partitioned, unrolled and pipelined to optimize power and area (NAND2 equivalent gate count) by using SystemC \textit{\#pragma} directives rather than having to hand code the partitioning, unrolling and pipelining in Verilog.

To increase the throughput of the \gls{pas} phase of the weight-shared-with-\gls{pasm} \gls{cnn} accelerator, the $imageBin$ array of line $12$ in Figure \ref{fig:mcmkPasmConvCode} is partitioned completely using the directive \textit{ARRAY\_PARTITION dim=1} (see line $2$) to inform Xilinx Vivado\_HLS to implement all bins in registers. When the \textit{for} loop of line $9$ to line $13$ is unrolled using the directive \textit{UNROLL} (see line $10$) and loop merged using the directive \textit{LOOP\_MERGE} (see line $11$), Vivado\_HLS implements $imageBin$ in registers rather than \gls{bram}, allowing the \gls{hls} to create multiple copies of the loop body so that it can parallelize the accumulation registers and associated accumulator logic and thus reduce the number of clock cycles of reads and writes to the $imageBin$ registers. 

The rest of the loops including the post pass \gls{mac} loop on lines $33$ and $42$ are pipelined with the directive \textit{PIPELINE II=1 rewind} which has an iteration interval of 1, suggesting to Vivado\_HLS that the loops shall need to process a new input every cycle which Vivado\_HLS will try to meet if possible. The \textit{rewind} option is used with the pipeline function to enable continuous loop pipelining such that there is no pause between one loop iteration ending and the next beginning. This is effective as there are perfect nested loops in the convolution. 

The partitioning, unrolling and loop merging reduces the latency cycles of the non-weight-shared, weight-shared and weight-shared-with-\gls{pasm} accelerators by $92\%$ at the expense of increasing the increasing the flip flop count by $97\%$ and thus the power and area of these combined function and loop pipeline registers. Implementing the $imageBin$ array in registers allows for cell compatibility in the \gls{asic} synthesis tool and quick synthesis time as no \gls{sram} needs to be modelled and implemented to store the input $image$ and $output feature$ values. This increased power and area overhead of the accelerators is a good trade-off for the increased throughput and lower latency.

The three versions of the \gls{cnn} accelerators are based on the AlexNet \cite{ImageNetClassificationWithDeepConvolutionalNeuralNetworks2012:Krizhevsky} \gls{cnn} and accelerate one layer of the convolution to allow for implementation in an \gls{fpga}. The accelerators include stride, an activation function, \gls{relu}, and bias (a means for the network to learn more easily) as the activation function and bias parameters are not shared. Striding (lines $4$, $5$ and $42$ of Figure \ref{fig:mcmkPasmConvCode}) allows for compression of the image or input feature map by allowing differing pixel strides of the kernel across the input feature map. For a stride value of 1, the kernel is moved across the input feature map at a stride of one pixel at a time. With a stride of 2 or more the kernel jumps 2 or more pixels as the kernel strides across the feature map. This sliding of the kernels produces smaller spatial output feature maps. The use of \gls{pasm} in the weight-shared accelerator is transparent to the functionality of the stride, activation function or biasing. Note that the numbers of weight parameters for a weight-shared system must be clustered (usually with K-means) and quantized to fit into 16 - 256 bins (see Han \MakeLowercase{\textit{et al's.}} \citeyear{DeepCompression2015:Han,eie2016:Han} research) as this reduction in numbers of weights is what allows \glspl{pasm} reduction in power and area by doing the \gls{pas} accumulations first followed by a single post pass \gls{mac}.

Our accelerators are high-level-synthesized to a hierarchical Verilog netlist using Xilinx Vivado\_HLS (version 2017.1) which allowed for quick functional simulation and hardware co-simulation and could also allow for implementation in both \gls{asic} and later \gls{fpga}. Vivado\_HLS reports the approximate latency of the design along with the approximate utilization results for \gls{bram}, \gls{dsp}, flip flops and \glspl{lut} after high level synthesis has been executed.

When implementing the accelerators in \gls{fpga}, Xilinx Vivado (version 2017.1) is used to synthesize, optimize, place and route the netlist from Xilinx Vivado\_HLS into a Xilinx 7-series Zynq XC7Z045 \gls{fpga} part running at 200MHz. When implementing the accelerators into a 45nm process ASIC running at 1GHz, Cadence Genus is used to synthesize and optimize the design for \gls{asic}. Cadence Genus supplies commands for reporting approximate timing, gate count and power consumption of the designs at the post-synthesis stage. The ``report timing'', ``report gates'' and ``report power'' commands of Cadence Genus are used to obtain the \gls{asic} timing, gate count and power results. The gate count is normalized to a NAND2 gate and this number reported as the overall gate count.

The designs are coded using integer/fixed point precision numbers (INTs). The bit widths of the image are maintained at 32-bit INTs whilst the weights are stored as variable 8-bit, 16-bit and 32-bit INTs. The bin indexes are stored as $2^2$-bits for 4 weights up to $2^4$-bits for 16 weights.

The encoding of finite state machines is set to gray encoding in order to keep the power consumption of the designs to a minimum. All registers and memories in the accelerators derived from variables in the SystemC are reset or initialized to zero. The resets are set as active low synchronous resets.

The number of kernels $M$ is kept small, i.e. $M=2$ to keep the synthesis time the \gls{asic} tools to a minimum. The number of channels $C$ is made as large as possible such that the $C \times KX \times KY$ is larger than $B$ bins to demonstrate the power saving effect of \gls{pasm} compared to the same number of channels for the weight-shared version of the accelerator, as suggested in Table \ref{table:complexity} and demonstrated in Table \ref{table:channelsKernelsMatrix}.

The $image$ cache was kept to a small tile of the image of multiple channels ($IH=5, IW=5, C=15$), to allow its implementation in a register file. However, the $image$ cache could be implemented in \gls{sram} in an \gls{asic}. This would allow for a larger cache storage of image and weight values and further reduce the power and area of the accelerators but would require more ``back-end'' layout design work of the accelerator, something not considered for this paper\footnote{The OSU FreePDK 45nm \gls{asic} process cell library used for the experiments does not have a facility to synthesize on-chip \gls{sram} in our implementation of the weight-shared-with-\gls{pasm} accelerator. If we had access to a library that would allow \gls{sram} synthesis we would be able to operate on larger data blocks in our \gls{asic} design. The weight-shared-with-\gls{pasm} is likely to be even more effective with larger input blocks (particularly a large value of $C$), because the cost of the post-pass multiplication can be amortized over more inputs.}. The $binIndex$ would remain in a register file as a maximum of $16 \times 32$ -bit values would be stored.

To further ensure the lowest number of multipliers utilized in the \gls{pasm} accelerator the \textit{allocation} directive is used to ensure that only one post pass multiplier is used further reducing the area and power whilst very slightly increasing the latency.

The Verilog netlists that are produced by Xilinx Vivado\_HLS are synthesized for \gls{asic} to produce a gate level netlist. Timing constraints in \gls{sdc} are created \cite{FpgaXdcTiming2015:Gangadharan} so all versions of the accelerator meet timing at 1GHz with a short 0.01ns clock transition using Cadence Genus (version 17.11) synthesizer.

The synthesis targets the OSU FreePDK 45nm \gls{asic} process cell library. Timing, latency, gate count (normalized to a NAND2 gate) and power consumption at different $B$ bins and $W$ bit widths are captured. These values are approximations as they are the post-synthesis estimates. The values will be optimized when implemented in \gls{asic} or \gls{fpga}.

The weight-shared-with-\gls{pasm} introduces a delay in processing the output of the \gls{pas} units. The \gls{pas} unit has a throughput of one pair of inputs per cycle, and so computes the initial accumulated values in about $N$ cycles, where:
\[N=(KX \times KY) \times C\]
The post pass \gls{mac} unit also has a throughput of one pair of inputs per cycle, so requires one cycle for each of the $B$ accumulator bins, for a total of $N+B$ \gls{pasm} cycles. In contrast, a simple \gls{mac} unit requires just $N$ cycles, however, consumes significantly more area and power, when compared to an accelerator with more than one \gls{pas} per \gls{mac}. 

Table \ref{table:channelsKernelsMatrix} shows the number of \gls{mac} operations that contribute to each output for various values of $C$ and $KX$ and $KY$. For example, if $C=32$ input channels are used with kernels of dimensions $KX \times KY=5 \times 5$, then each computed value will be the result of 800 \gls{mac} operations. A simple fully-pipelined \gls{mac} unit might be able to compute this result in a little more than 800 cycles. As can be seen from lines $11$ to $13$ of Figure \ref{fig:mcmkConvCode}, each element of the output of a convolution layer of a \gls{cnn} is the result of $C \times KX \times KY$ multiply-accumulate operations, or 800 cycles in this example.

In contrast, a \gls{pasm} has two phases: a \gls{pas} phase, and a post-pass \gls{mac} phase. The \gls{pas} phase computes a histogram of the frequency of each weight input, and depends entirely on the number of inputs. However, the post-pass \gls{mac} phase depends not on the number of inputs, but on the number of different weights that can appear (each of which occupies one of the $B$ bins). Provided the number of inputs, $C \times KX \times KY$, is much larger than the number of bins, $B$, the cost of the post-pass remains small relative to the cost of the \gls{pas} phase.  For example, if $B=16$, then the cost of the post-pass will be a small fraction of the $800$ operations needed at the \gls{pas} phase. Careful consideration of the size of bins used with respect to the number of channels and kernels is important due to the $summands$ being multiply-accumulated many times before the $outFeat$ is updated as can be seen on lines 11 - 13 of Figure \ref{fig:mcmkConvCode}. The number of accumulations should therefore be much larger than $B$ for \gls{pasm} to be efficient in a weight-shared convolution accelerator.

\begin{table}[H]
	\centering
	\caption{Typical Numbers of \gls{mac} Operations}
	\begin{tabular}{cllll}
		\multicolumn{1}{l}{}                  &                                   & \multicolumn{3}{c}{\textbf{input\_channels ($C$)}}                                                                \\ \cline{3-5} 
		\multirow{5}{*}{\textbf{kernels ($K$)}} & \multicolumn{1}{l|}{\textbf{}}    & \multicolumn{1}{l|}{\textbf{32}} & \multicolumn{1}{l|}{\textbf{128}} & \multicolumn{1}{l|}{\textbf{512}} \\ \cline{2-5} 
		& \multicolumn{1}{l|}{\textbf{1x1}} & \multicolumn{1}{l|}{32}          & \multicolumn{1}{l|}{128}          & \multicolumn{1}{l|}{512}          \\ \cline{2-5} 
		& \multicolumn{1}{l|}{\textbf{3x3}} & \multicolumn{1}{l|}{288}         & \multicolumn{1}{l|}{1152}         & \multicolumn{1}{l|}{4608}         \\ \cline{2-5} 
		& \multicolumn{1}{l|}{\textbf{5x5}} & \multicolumn{1}{l|}{800}         & \multicolumn{1}{l|}{3200}         & \multicolumn{1}{l|}{12800}        \\ \cline{2-5} 
		& \multicolumn{1}{l|}{\textbf{7x7}} & \multicolumn{1}{l|}{1568}        & \multicolumn{1}{l|}{6272}         & \multicolumn{1}{l|}{25088}        \\ \cline{2-5} 
	\end{tabular}
	\label{table:channelsKernelsMatrix}
\end{table}

\begin{figure}
	\centering
	\includegraphics[width=0.6\linewidth]{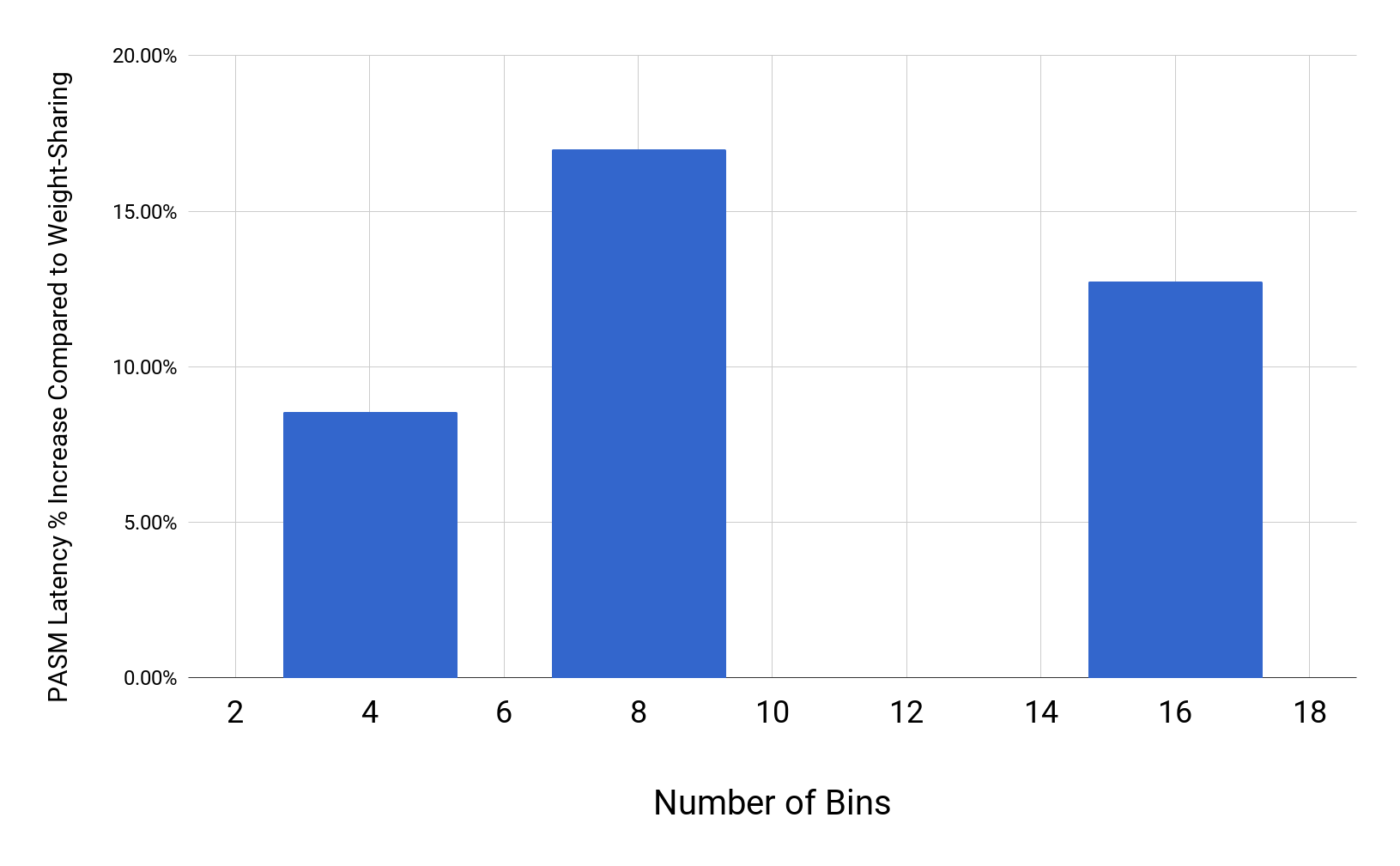}
	\caption{Latency of weight-shared-with-\gls{pasm} convolution compared to weight-shared convolution}
	\label{fig:latencyComparison}
\end{figure}

{\section{Evaluation of PASM in a CNN accelerator}
	\label{sec:evaluation}}
{\subsection{ASIC Results}
	\label{subsec:asicResults}}
The \gls{pasm} is implemented in a weight-shared \gls{cnn} accelerator and synthesized into an \gls{asic}. The latency is compared with that of the weight-shared accelerator. The latency results for each of the non-weight-shared, weight-shared and weight-shared-with-\gls{pasm} accelerators is obtained from Vivado\_HLS Synthesis reports and the percentage differences graphed as seen in Figure \ref{fig:latencyComparison}. The latency of the weight-shared-with-\gls{pasm} in Figure \ref{fig:latencyComparison} was between 8.5\% for 4-bin and 12.75\% for 16-bin greater than that of the corresponding weight-shared version, which is expected due to the indirection of the \gls{pas} units.

Latency can be further reduced by relaxing the \textit{ALLOCATION} directive (see line 3 of Figure \ref{fig:mcmkPasmConvCode}) constraint on the multiplier. If more post-pass multipliers are used then the latency drops with a corresponding increase in power and area which may be acceptable depending on target device resources available.

\begin{figure}
	\centering
	\subcaptionbox{\gls{asic} Gate Count for 32 Bit Kernel, 4-bin Accelerators\label{fig:asic4BinUtilizationComparison}}{\includegraphics[width=0.48\linewidth]{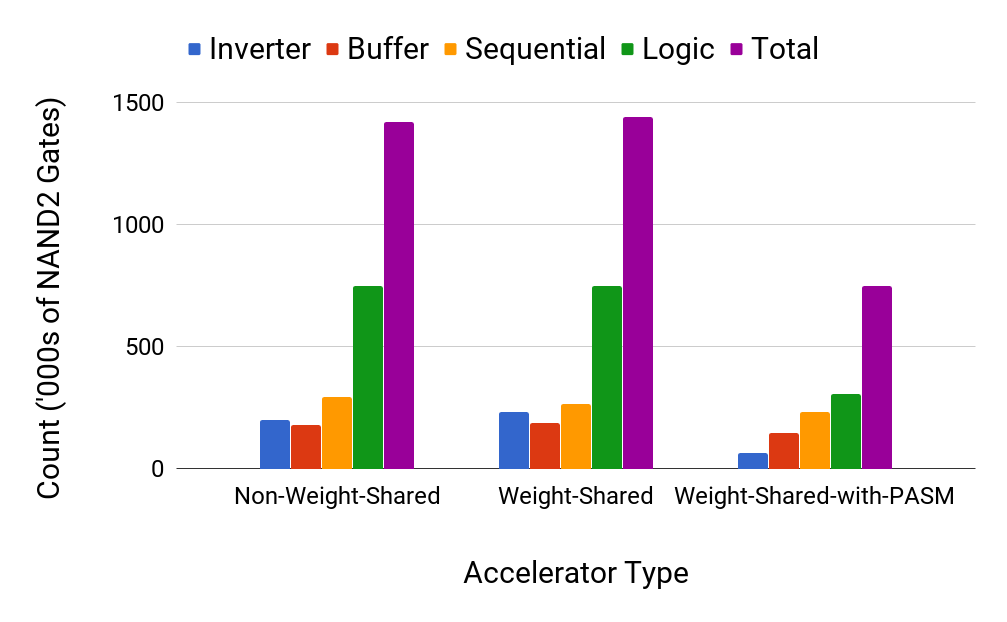}}
	\subcaptionbox{\gls{asic} Power Consumption for 32 Bit Kernel, 4-bin Accelerators\label{fig:asic4BinTotalDynamicPowerComparison}}{\includegraphics[width=0.48\linewidth]{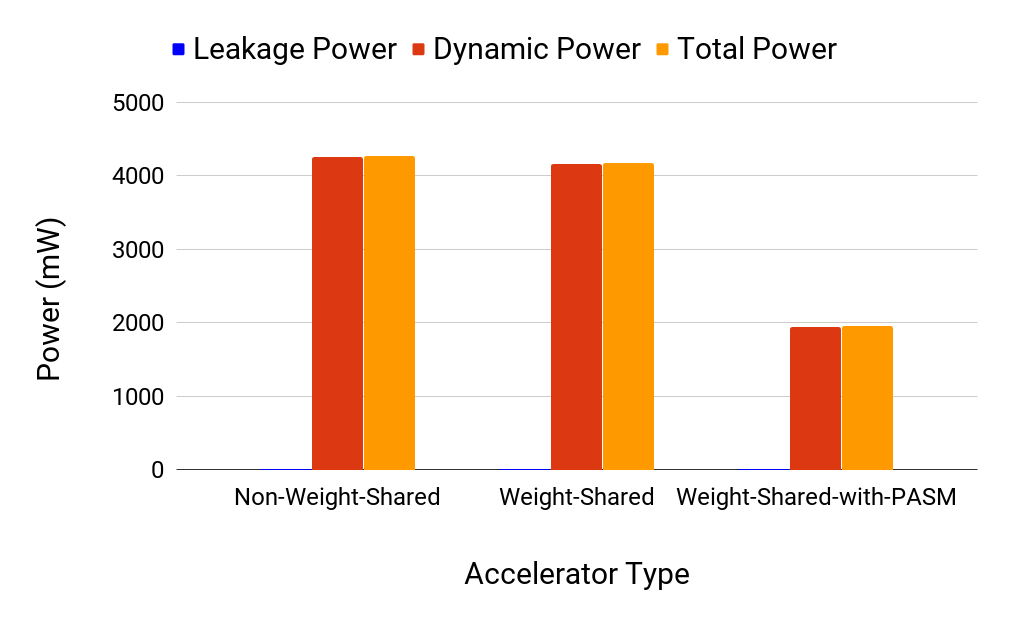}}
	\caption{4-bin, 32-bit Kernel weight-shared-with-PASM vs Weight Shared Gate Count and Power Comparisons in \gls{asic}}
\end{figure}

For a 4-bin \gls{pasm} accelerator, with 32-bit wide kernels, Figure \ref{fig:asic4BinUtilizationComparison} shows the gate count reports obtained from Cadence ``report gates'' command and normalized to a NAND2 gate. \gls{pasm} uses 47.2\% fewer total NAND2 gates compared with the non-weight-shared version and 47.8\% fewer total NAND2 gates compared with weight-shared design. Figure \ref{fig:asic4BinTotalDynamicPowerComparison} obtained from Cadence ``report power'' command, \gls{pasm} uses 54.3\% less total power when compared with its non-weight-sharing counterpart and 53.2\% less total power when compared with the weight-shared version.

\begin{figure}
	\centering
	\subcaptionbox{\gls{asic} Gate Count for 32 Bit Kernel, 8-bin Accelerators\label{fig:asic8BinUtilizationComparison}}{\includegraphics[width=0.48\linewidth]{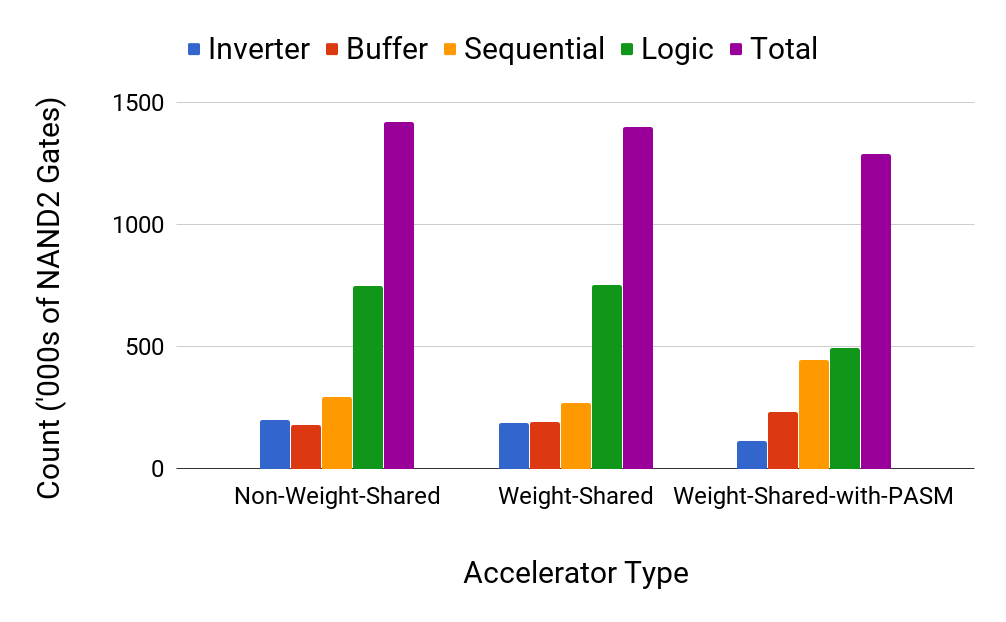}}
	\subcaptionbox{\gls{asic} Power Consumption for 32 Bit Kernel, 8-bin Accelerators\label{fig:asic8BinTotalDynamicPowerComparison}}{\includegraphics[width=0.48\linewidth]{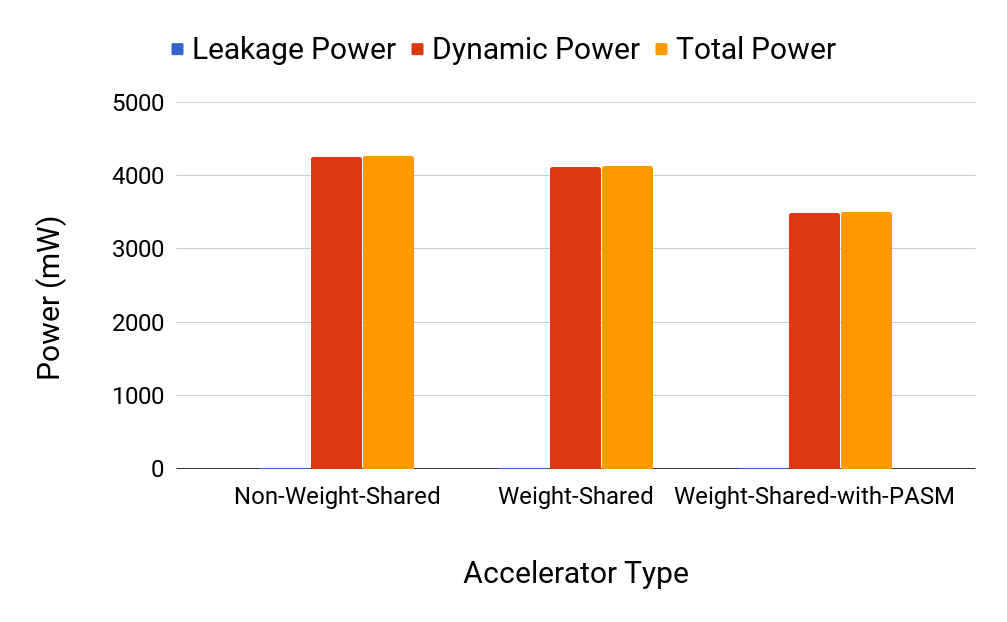}}
	\caption{8-bin, 32-bit Kernel weight-shared-with-PASM vs Weight Shared Gate Count and Power Comparisons in \gls{asic}}
\end{figure}

For an 8-bin, 32-bit wide kernel \gls{pasm} accelerator, Figure \ref{fig:asic8BinUtilizationComparison} obtained from Cadence ``report gates'' command and normalized to a NAND2 gate, \gls{pasm} uses 9.4\% fewer total NAND2 gates compared with the non-weight-shared and 8.1\% fewer total NAND2 gates compared with the weight-shared accelerators. Figure \ref{fig:asic8BinTotalDynamicPowerComparison} obtained from Cadence ``report power'' command, \gls{pasm} consumes 18.1\% less total power when compared with its non-weight-sharing and 15.2\% less total power when compared with the weight-sharing accelerator. 

\begin{figure}
	\centering
	\subcaptionbox{\gls{asic} Gate Count for 32 Bit Kernel, 16-bin Accelerators\label{fig:asic16BinUtilizationComparison}}{\includegraphics[width=0.48\linewidth]{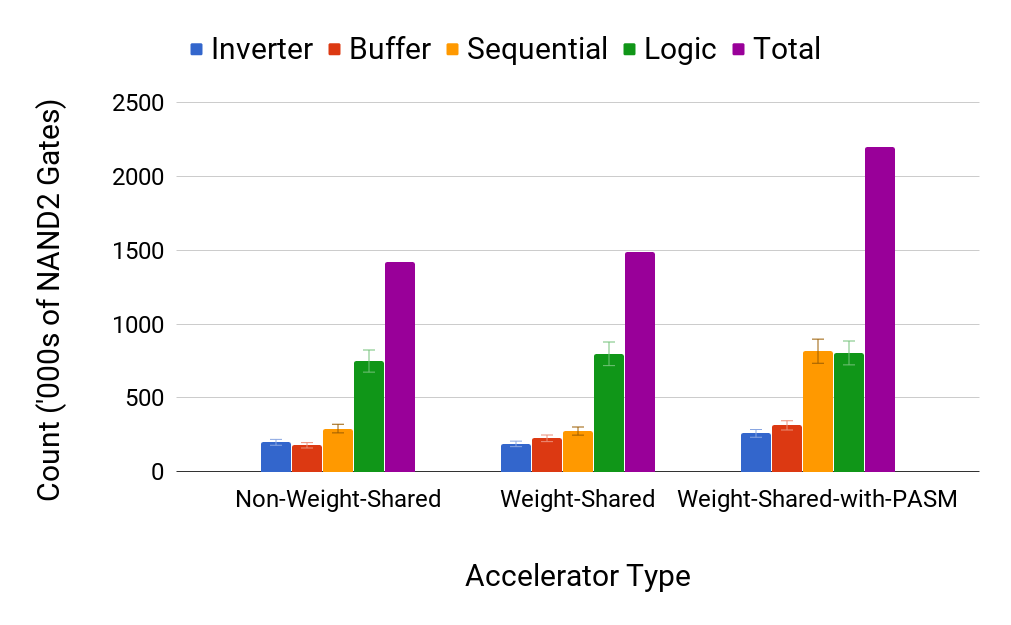}}
	\subcaptionbox{\gls{asic} Power Consumption for 32 Bit Kernel, 16-bin Accelerators\label{fig:asic16BinTotalDynamicPowerComparison}}{\includegraphics[width=0.48\linewidth]{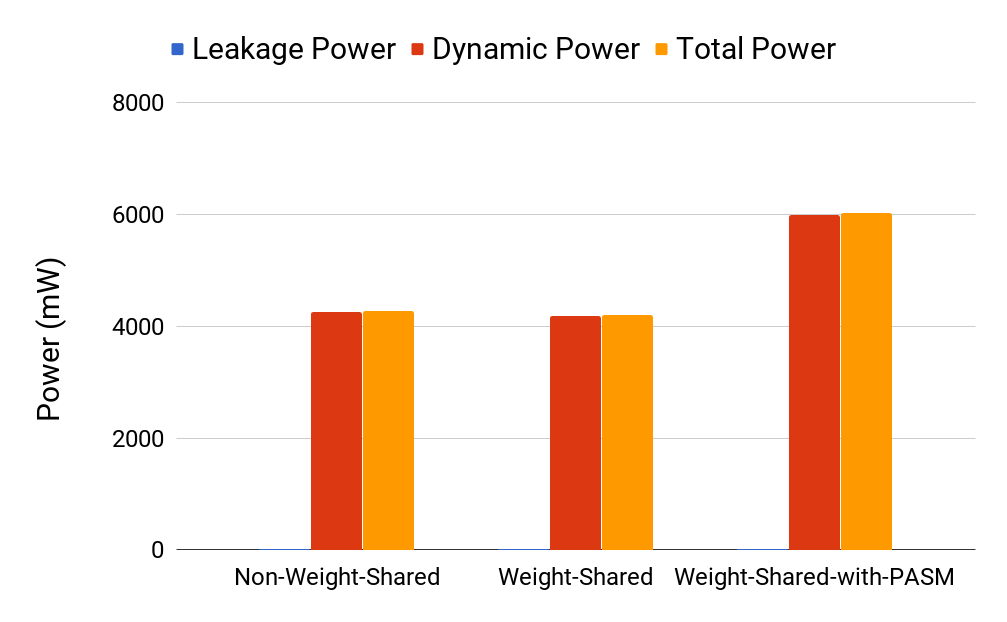}}
	\caption{16-bin, 32-bit Kernel weight-shared-with-PASM vs Weight Shared Gate Count and Power Comparisons in \gls{asic}}
\end{figure}

For a 16-bin, 32-bit wide weight-shared-with-\gls{pasm} accelerator, \gls{pasm} no longer offers a good return with this level of unrolling, pipelining and partitioning of the \textbf{\textit{imageBin}}, at least when targeted at a 1GHz \gls{asic} with this 45nm process cell library as it uses more NAND2 gates (see Figure \ref{fig:asic16BinUtilizationComparison}) and power (see Figure \ref{fig:asic16BinTotalDynamicPowerComparison}) compared with the weight-shared accelerator. This is due to the \gls{asic} tools having to increase the area and therefore power to meet timing at 1GHz for the 16-bins at 32-bits wide \gls{pasm}. To achieve better power and area results for \gls{pasm} at 16-bins or greater, it might be better to target a lower clock frequency, for example 800MHz. Alternatively, use a more efficient geometry \gls{asic} cell library. Design changes could be made to reduce pipelining and unrolling of the levels of the inner four of the \textit{for} loops of the convolutional code, which would reduce the area and power whilst making it easier for the \gls{asic} tools to achieve timing, however, this may increase latency of the accelerator.

\begin{figure}[t]
	\centering
	\subcaptionbox{\gls{asic} Gate Count for 8 Bit Kernel, 4-bin Accelerators\label{fig:asic4BinInt8UtilizationComparison}}{\includegraphics[width=0.48\linewidth]{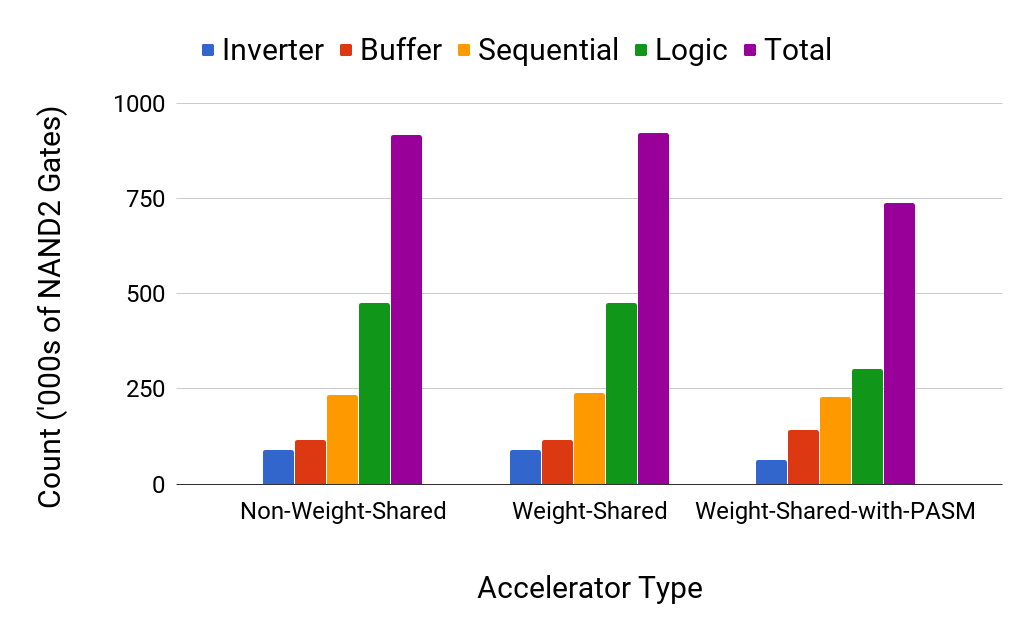}}
	\subcaptionbox{\gls{asic} Power Consumption for 8 Bit Kernel, 4-bin Accelerators\label{fig:asic4BinInt8WideTotalPowerComparison}}{\includegraphics[width=0.48\linewidth]{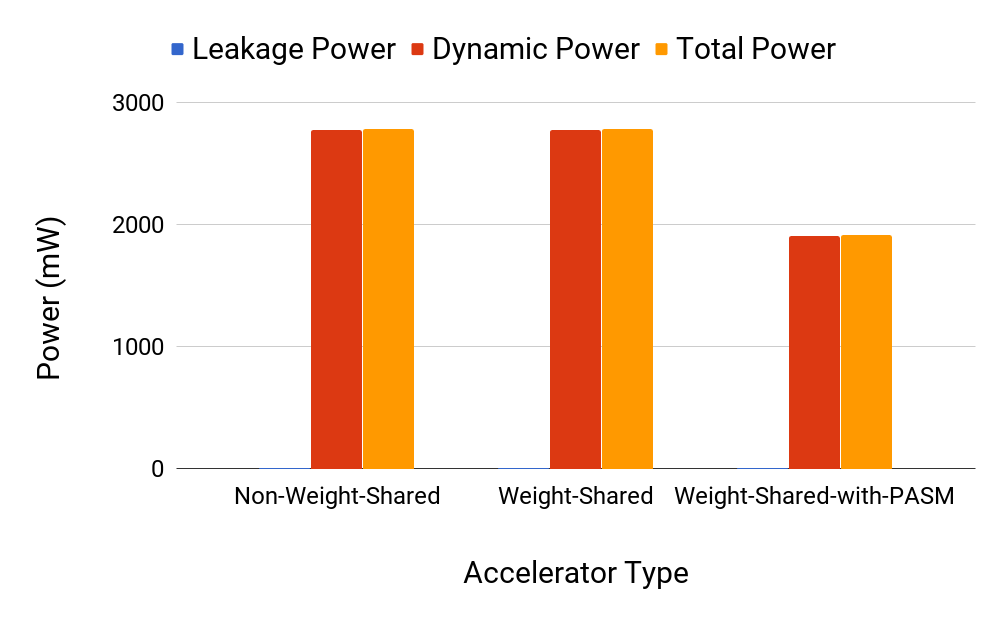}}
	\caption{4-bin, 8 bit Kernel weight-shared-with-PASM vs Weight Shared Gate Count and Power Comparisons in \gls{asic}}
	\vspace{-0.6cm}
\end{figure}

Due to the increased academic and industrial interest in applying INT8 approximations to reduce memory storage and bandwidth of the kernel data \cite{8BitApproxForDL2015:Dettmers,DeepLearningWithInt82017:Fu}, we show the results for the 8-bit kernel versions of the accelerators with 4-bins. This demonstrates that for a bin depth of $4$, \gls{pasm} achieves a 19.8\% reduction in gate count, Figure \ref{fig:asic4BinInt8UtilizationComparison} obtained from Cadence ``report gates'' command and normalized to a NAND2 gate and a 31.3\% reduction in power compared to the weight-sharing version, Figure \ref{fig:asic4BinInt8WideTotalPowerComparison} obtained from Cadence ``report power'' command.

\vspace{0.4cm}
{\subsection{FPGA Results}
	\label{subsec:fpgaResults}}
We implement the weight-shared-with-\gls{pasm} accelerator in the Xilinx 7-series Zynq \gls{fpga}, the XC7Z045 part implemented on the Zynq ZC706 development board. Timing constraints in \gls{xdc} are created such that the accelerator designs met timing at 200MHz. The resets are set as active high synchronous resets for better \gls{fpga} power performance. The state machines are set to gray encoding.

The \textbf{\textit{image}}, \textbf{\textit{imageBin}}, and \textbf{\textit{kernel}} were cached in \gls{bram} in the \gls{fpga}. This allows for a larger cache storage of image and weight values and further reduce the power and area of the accelerator. However, a larger image and kernel cache could be employed for greater throughput of the accelerators but for the purposes of comparison with the \gls{asic} implementation, the same image and kernel dimensions are used.

When using the \textit{UNROLL} and \textit{PIPELINE} directives with the \textit{for} loops and using Vivado\_HLS synthesis followed by \gls{rtl} synthesizing and fully implementing the designs with Vivado, the non-weight shared and weight-shared versions of the 16-bin, 32-bit kernel data designs utilizes 405 \gls{dsp} units on the \gls{fpga} of the ZC706 board. If a smaller, more resource constrained \gls{fpga} is required for cost reasons, like the Xilinx XC7Z020 part found on the Xilinx PYNQ-Z1 low cost development board, then the non-weight shared and weight-shared versions of the design would over utilize the 220 \gls{dsp} units of the PYNQ-Z1 board's XC7Z020 \gls{fpga} part.

\begin{figure}
	\centering
	\subcaptionbox{Cell Count for 32 Bit Kernel, 4-bin Accelerators\label{fig:fpga4BinUtilizationComparison}}{\includegraphics[width=0.48\linewidth]{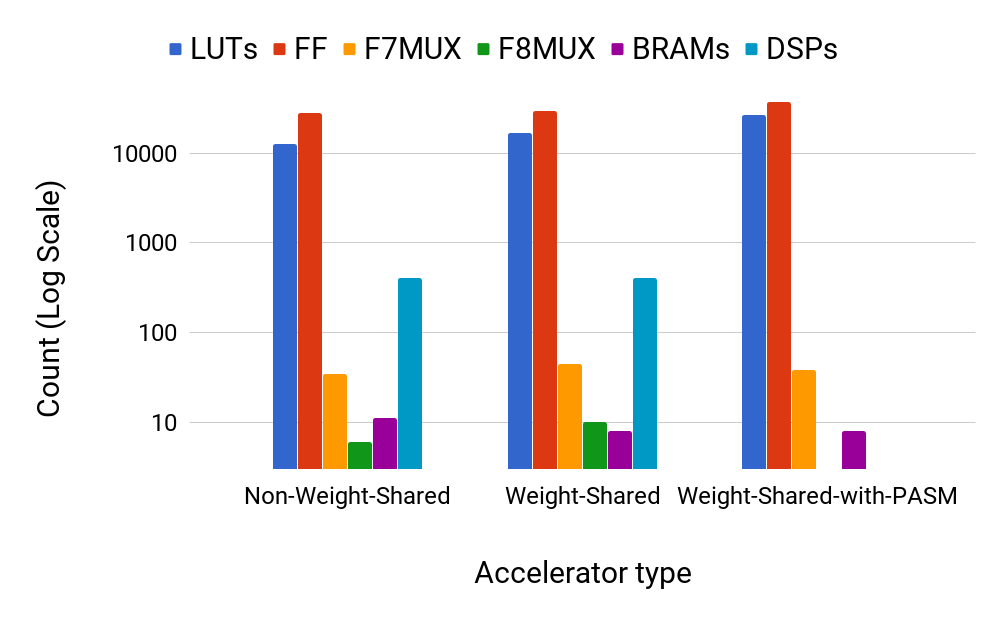}}
	\subcaptionbox{Power Consumption for 32 Bit Kernel, 4-bin Accelerators\label{fig:fpga4BinTotalDynamicPowerComparison}}{\includegraphics[width=0.48\linewidth]{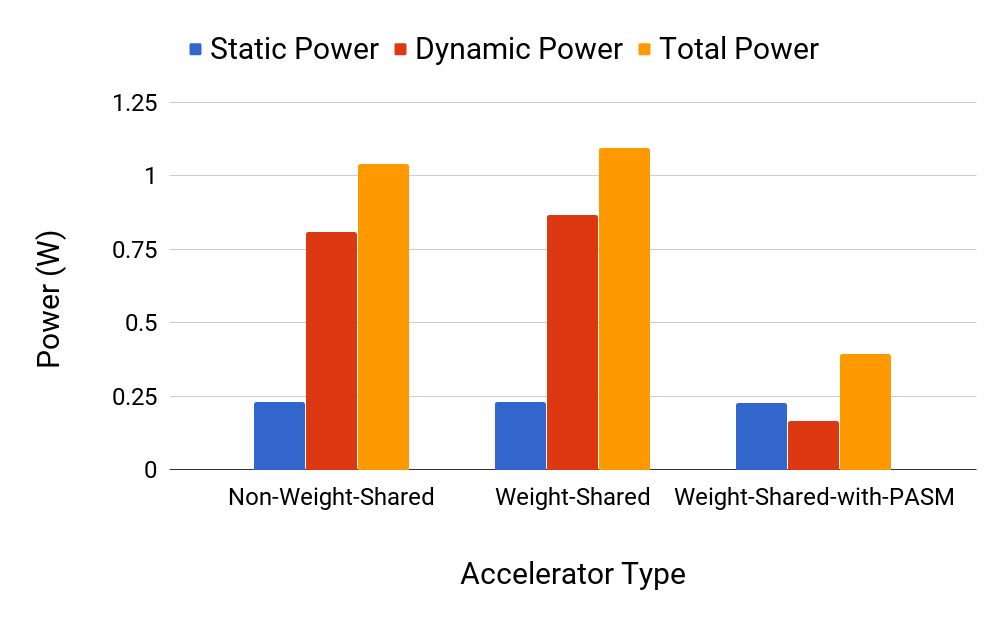}}
	\caption{4-bin 32-bit Kernel weight-shared-with-PASM vs Weight Shared Gate Count and Power Comparisons in \gls{fpga}}
\end{figure}

The weight-shared-with-\gls{pasm} version of the design for the same 4-bin, 32-bit kernel, Figure \ref{fig:fpga4BinUtilizationComparison} obtained with Vivado's ``report\_utilization'' command, similarly unrolled and pipelined, \gls{hls} synthesized in Vivado\_HLS followed by \gls{rtl} synthesized and fully implemented in Vivado, only utilizes 3 \gls{dsp} units, 99\% fewer \glspl{dsp} than the other versions of the accelerator with the same 12\% increase in latency as the \gls{asic} implementation. \gls{pasm} also consumes 28\% fewer \glspl{bram}, whilst consuming 64\% less power than the weight-shared accelerator, Figure \ref{fig:fpga4BinTotalDynamicPowerComparison}, obtained with Vivado's ``report\_power'' command. Increasing the number of post-pass \glspl{mac} decreases the latency slightly whilst increasing the power consumption and \gls{dsp} usage, as the bottleneck is in the accumulators of the \gls{pas} which can be seen as the number of $C \times KX \times KY$ accumulations which must be larger than that of $B$ bins for \gls{pasm} to be effective and efficient as seen in Table \ref{table:complexity} and Table \ref{table:channelsKernelsMatrix}.

\begin{figure}
	\centering
	\subcaptionbox{Cell Count for 32 Bit Kernel, 8-bin Accelerators\label{fig:fpga8BinUtilizationComparison}}{\includegraphics[width=0.48\linewidth]{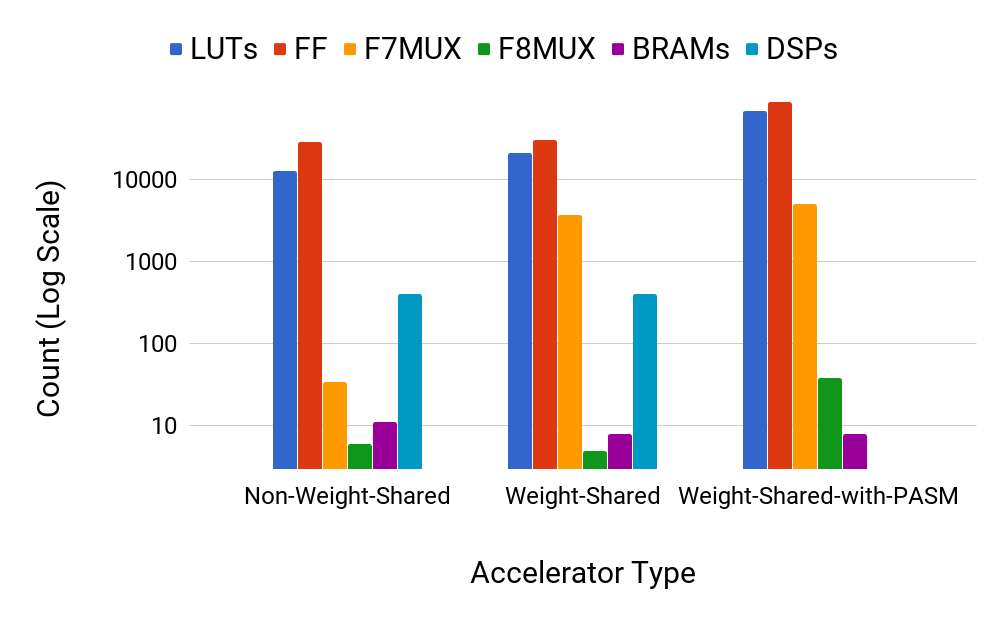}}
	\subcaptionbox{Power Consumption for 32 Bit Kernel, 8-bin Accelerators\label{fig:fpga8BinTotalDynamicPowerComparison}}{\includegraphics[width=0.48\linewidth]{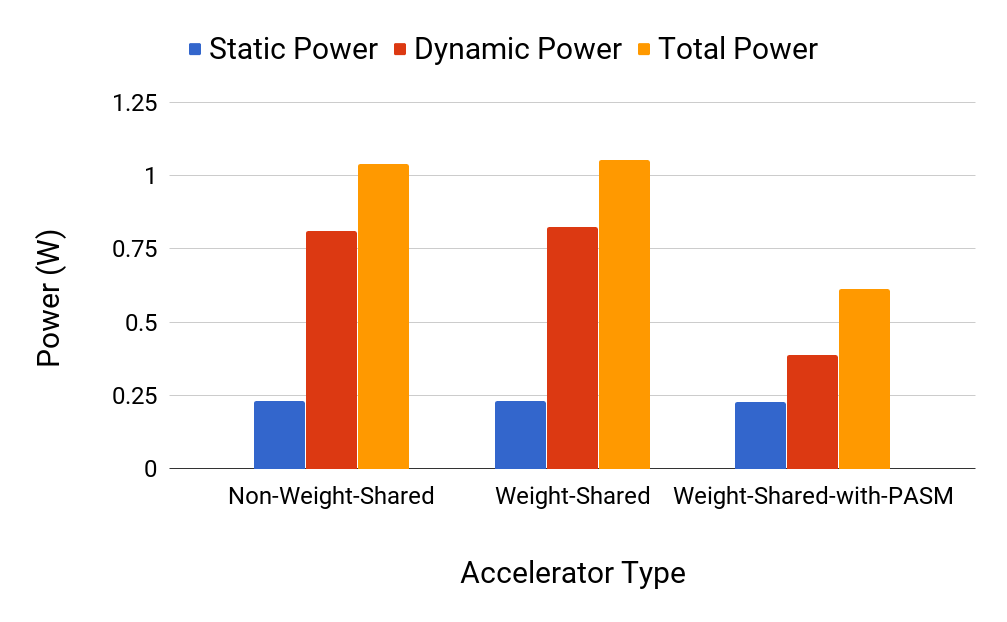}}
	\caption{8-bin 32-bit Kernel weight-shared-with-PASM vs Weight Shared Gate Count and Power Comparisons in \gls{fpga}}
\end{figure}

For an 8-bin \gls{pasm} accelerator, with 32-bit kernels, Figure \ref{fig:fpga8BinUtilizationComparison} obtained with Vivado's ``report\_utilization'' command, \gls{pasm} uses 99\% fewer \glspl{dsp} and 28\% fewer \glspl{bram} compared with the weight-shared design. Figure \ref{fig:fpga8BinTotalDynamicPowerComparison} obtained with Vivado's ``report\_power'' command, \gls{pasm} uses 41.6\% less total power when compared with its weight-shared version.

\begin{figure}
	\centering
	\subcaptionbox{Cell Count for 32 Bit Kernel, 16-bin Accelerators\label{fig:fpga16BinUtilizationComparison}}{\includegraphics[width=0.48\linewidth]{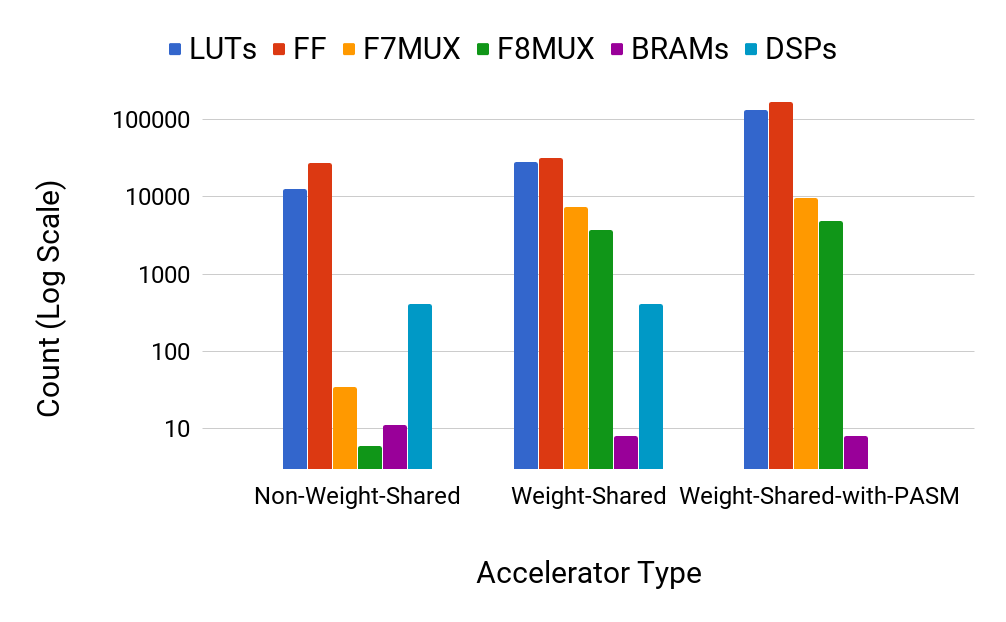}}
	\subcaptionbox{Power Consumption for 32 Bit Kernel, 16-bin Accelerators\label{fig:fpga16BinTotalDynamicPowerComparison}}{\includegraphics[width=0.48\linewidth]{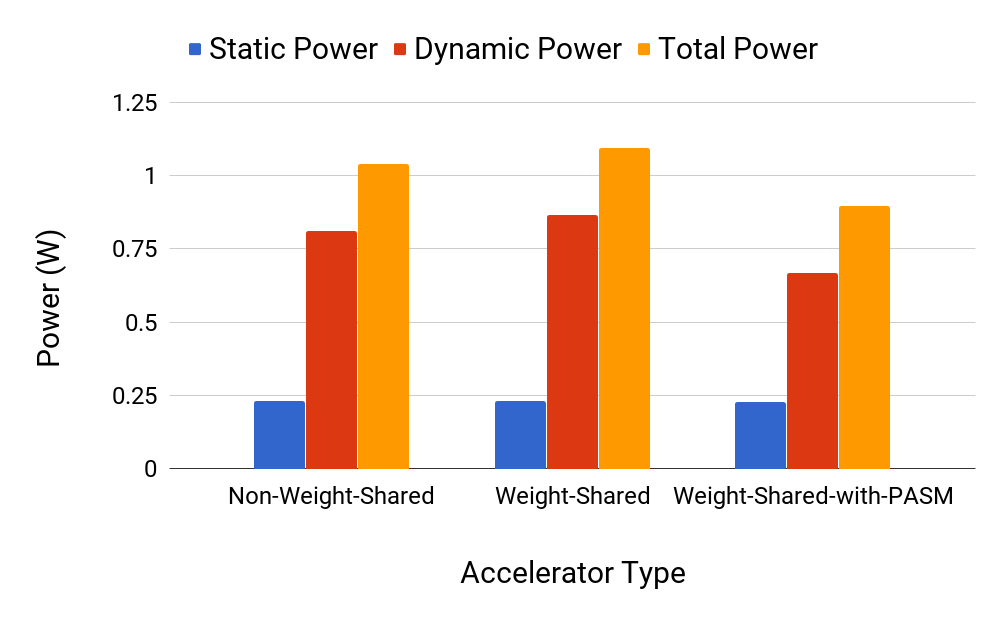}}
	\caption{16-bin 32-bit Kernel weight-shared-with-PASM vs Weight Shared Gate Count and Power Comparisons in \gls{fpga}}
\end{figure}

For a 16-bin \gls{pasm} accelerator, with 32-bit kernels the utilization reported with Vivado's ``report\_utilization'' command, Figure \ref{fig:fpga16BinUtilizationComparison}, \gls{pasm} uses 99\% fewer \glspl{dsp} and 28\% fewer \glspl{bram} compared with the weight-shared design. Figure \ref{fig:fpga16BinTotalDynamicPowerComparison}, \gls{pasm} uses 18\% less total power when compared with its weight-shared version, reported with Vivado's ``report\_power'' command.

It is also possible to clock the \gls{pasm} at higher clock speeds for the same latency than that of the weight-shared counterpart but again for the sake of comparison, clock speeds are kept consistent between all versions of the accelerators.

\begin{figure}
	\centering
	\subcaptionbox{Cell Count for 8 Bit Kernel, 8-bin Accelerators\label{fig:fpga8BinInt8UtilizationComparison}}{\includegraphics[width=0.48\linewidth]{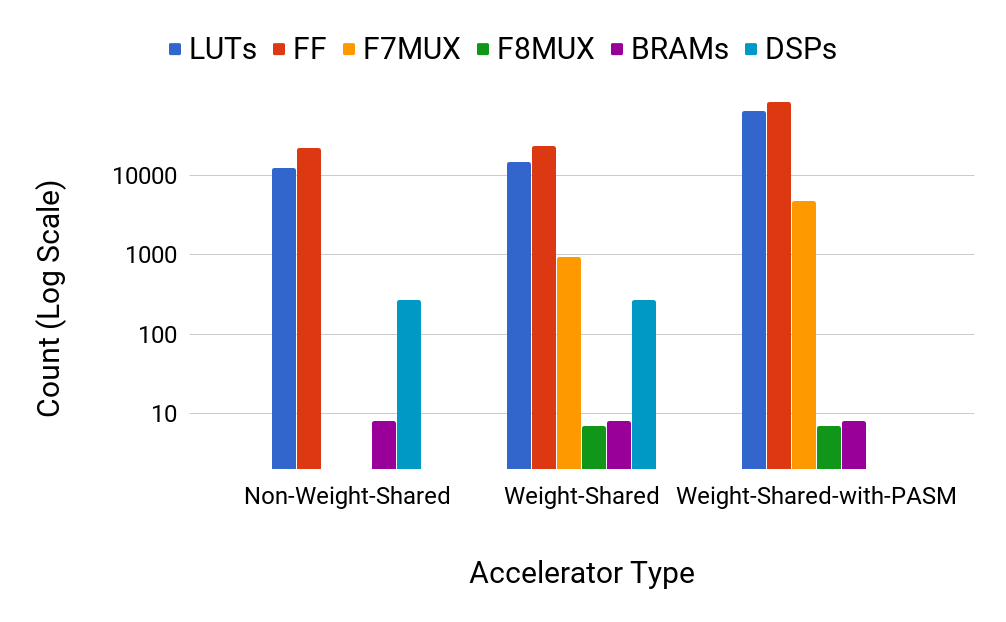}}
	\subcaptionbox{Power Consumption for 8 Bit Kernel, 8-bin Accelerators\label{fig:fpga8BinInt8TotalDynamicPowerComparison}}{\includegraphics[width=0.48\linewidth]{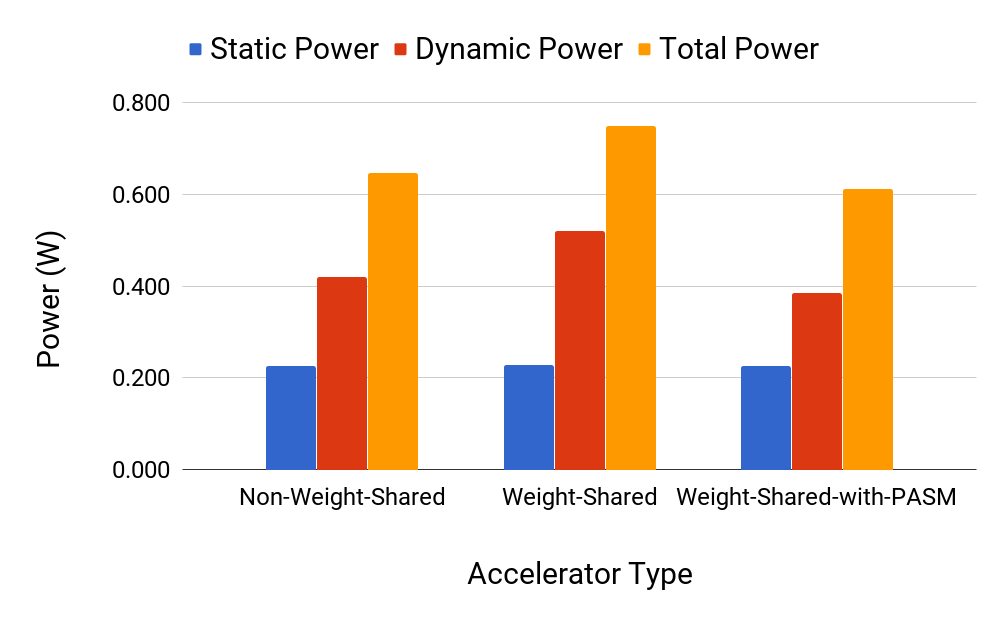}}
	\caption{8-bin 8 bit Kernel weight-shared-with-PASM vs Weight Shared Gate Count and Power Comparisons in FPGA}
\end{figure}

If INT8 approximations are desired for the weight data, an 8-bit wide, 8-bin \gls{pasm} accelerator, Figure \ref{fig:fpga8BinInt8UtilizationComparison} again obtained with Vivado's ``report\_utilization'' command, uses 99\% fewer \glspl{dsp} but the same number of \glspl{bram} as it's weight-shared counterpart. Figure \ref{fig:fpga8BinInt8TotalDynamicPowerComparison}, \gls{pasm} uses 18.3\% less total power when compared with its weight-shared version.

For a 16-bin, 8-bit wide \gls{pasm} accelerator, \gls{pasm} no longer offers a good return when targeted at a 200MHz \gls{fpga} with this level of unrolling, pipelining and partitioning of the \textbf{\textit{imageBin}} as it uses more flip-flop gates and power, exceeding the gate count and power of the \gls{dsp} units being used in the weight-shared accelerator. At this stage, it would be better to either implement the \textbf{\textit{imageBin}} in dual port \gls{bram} and incur a slight increase in latency or do not unroll and pipeline as many levels of the inner four of the \textit{for} loops of the convolutional code.

{\subsection{Overall Results}
	\label{subsec:overallResults}}
The precision of the results of a weight-shared \gls{cnn} accelerator that uses \gls{pasm} are identical to that of a weight-shared \gls{cnn} accelerator using traditional \glspl{mac}. The same filters and image data are being used for the weight-shared accelerator as demonstrated in Figure \ref{fig:weightSharedMac} and the weight-shared-with-\gls{pasm} accelerator shown in Figure \ref{fig:pasmInOperation}. Whilst \gls{pasm} has a different underlying process of permuting the convolution, the results of a convolution layer are identical to that of a standard \gls{mac} weight-shared accelerator, except \gls{pasm} adds a 12.5\% increase in latency in obtaining the result but with vastly reduced power consumption and area (NAND2 gates) compared to the traditional \gls{mac} version.

As suggested in Han \MakeLowercase{\textit{et al.}} \citeyear{eie2016:Han}, they show that the Top-5 classification accuracy of their weight-shared \gls{cnn} accelerator is 19.70\% compared to 19.73\% Top-5 accuracy of the baseline non-weight-shared \gls{cnn} accelerator due to there being many less filter weight values. When \gls{pasm} is used in a weight-shared \gls{cnn} accelerator the classification accuracy is unaffected when compared to the baseline weight-shared \gls{cnn} accelerator counterpart as the same filter weight values of the weight-shared \gls{cnn} accelerator are used and the same output feature map results are obtained.

\gls{pasm} is beneficial for up to 16 weight bins and 32-bits for \gls{fpga} at 200MHz and 8 weight bins and 32-bits for \gls{asic} at 1GHz 45nm process when coded using SystemC with the above unrolling, pipelining and partitioning configuration. As demonstrated earlier in the paper, were a weight-shared-with-\gls{pasm} \gls{cnn} accelerator to be coded in Verilog, the numbers of bins supported could indeed be higher. We wanted to experiment with differing pipelining, unrolling and partitioning directives and their effect on making \gls{pasm} more efficient, something which would have been impractical had it been coded in Verilog, so SystemC was used. Further SystemC and other \gls{sram} optimizations (for image and output feature map caching) could have been done to the accelerators, but this was not the focus of this paper and may be undertaken as future work.


\vspace{0.4cm}
{\section{Related Work}
	\label{sec:relatedWork}}
There have been many different \gls{cnn} hardware accelerators proposed for both \gls{fpga} and \gls{asic}. Gupta \MakeLowercase{\textit{et al.}} \citeyear{limitedNumericalPrecision2015:Gupta} show increased efficiency in an \gls{fpga} hardware accelerator of a 16-bit fixed-point representation using stochastic rounding without loss of accuracy. Zhang \MakeLowercase{\textit{et al.}} \citeyear{OptimizingFpgaAccelForCNN2015:Zhang} deduced the best \gls{cnn} accelerator taking \gls{fpga} requirements into consideration and then implement the best on an \gls{fpga} to demonstrate high performance and throughput. Chen \MakeLowercase{\textit{et al.}} \citeyear{DianNao2014:Chen} design an \gls{asic} accelerator for large-scale \glspl{cnn} focusing on the impact of memory on the accelerator performance.

Han \MakeLowercase{\textit{et al.}} \citeyear{eie2016:Han} have proposed an Efficient Inference Engine which builds on their `Deep compression' \citeyear{DeepCompression2015:Han} work to perform inferences on the deeply compressed network to accelerate the weight-shared matrix-vector multiplication. This accelerates the classification task whilst saving energy when compared to \gls{cpu} or \gls{gpu} implementations. Given that one aspect of deep compression is quantizing and dictionary encoding weights, we believe that the use of our PASM units might further reduce resource and energy requirements.

Chen \MakeLowercase{\textit{et al.}} \citeyear{Eyeriss2016a:Chen} address the problem of data movement which consumes large amounts of bandwidth and energy in their \textit{Eyeriss} accelerator. They focus on data flow in the \gls{cnn} to minimize data movement by reusing weights within the hardware accelerator to improve locality. This was implemented in \gls{asic} and power and implementation results compared showing the effectiveness of weight reuse in saving power and increasing locality. Chen \MakeLowercase{\textit{et al.}} reduce the required memory bandwidth primarily be reusing data that is already on-chip rather through weight compression. However, the two approaches are mostly orthogonal, so our PASM approach could potentially work together with an \textit{Eyeriss} type accelerator.

Ma \MakeLowercase{\textit{et al.}} \citeyear{OptimizingLoopOperation2017a:Ma} present an in-depth analysis of convolution loop acceleration strategies by numerically characterizing the loop optimization techniques. They do this by looking at different levels of loop unrolling and loop tiling (subdividing the design into smaller blocks) and loop interchange (different ordering of the loops). They also consider latency and partial sum storage and how they can minimize both. They provide design guidelines for an efficient implementation of the accelerator to minimize latency, minimize partial sum storage, minimize both on-chip buffer accesses and off-chip memory accesses. For the four inner convolution loops, they show which loops to unroll (in this case all four loops), which to tile (loops 1 and 2 are buffered), and which to interchange (compute loop1 then loop 2 but it doesn't matter the order of loop 3 and loop 4). They implement the accelerator for a VGG-16 CNN model in an Arria-10 GX 1150 \gls{fpga} (3600 \glspl{dsp}, $18 \times 18$ 20kb \glspl{ram}) at 150MHz and coded in Verilog achieving 645.25 \gls{gops} of throughput and 47.97ms of latency per image. This work on loop ordering is complementary to our work on architecting a lower-resource \gls{mac} unit.

Several research groups have studied the effects of lower precision weight values for \glspl{cnn}. Reducing the data precision of weight data is an alternative method of quantizing that is different to the weight sharing of Han \MakeLowercase{\textit{et al.}} \citeyear{eie2016:Han}. Rather than selecting a set of quantized values guided by the values in the data, low-precision approaches simply quantize existing weights to the nearest low-precision value. A particularly popular data type is 8-bit integers.

Dettmers \citeyear{8BitApproxForDL2015:Dettmers} shows how 8-bit for data and model parallelism increases the performance of machine learning whilst maintaining accuracy on MNIST, CIFAR10 and ImageNet neural networks. The paper describes data parallelism across multiple \glspl{gpu} showing bandwidth and latency limitations on the \gls{pcie}. They show different ways of representing the mantissa and exponent in the available 8-bits. They show how the 32-bit value is compressed into the 8-bit value and decompressed. They show how representing the 8-bits using a dynamic tree data type is able to approximate random numbers better than other known data types but interestingly all approximation techniques (dynamic tree, linear quantization, 8-bit mantissa and static tree) work well in training. They investigate other sub 32-bit data types and show that model parallelism in conjunction with sub-batches works very well in networks and avoids the problem of large batch sizes for 1-bit quantization proposed by Seide \MakeLowercase{\textit{et al.}} \citeyear{1-bitStochasticGradientDescent2014:Seide}.

In the Xilinx white paper of Fu \MakeLowercase{\textit{et al.}}, \cite{DeepLearningWithInt82017:Fu}, Xilinx makes use of 18- and 27-bit multipliers hardware multipliers that are commonly found on Xilinx \glspl{fpga}. They use these multipliers to compute two 8-bit multiplications in parallel, giving better performance and efficiency than if each multiplier were to perform just one 8-bit multiplication per cycle. This approach is quite different to our proposal for a new type of \gls{mac} unit, and depends on the presence of existing hard-coded multipliers on the \gls{fpga}.


\vspace{0.4cm}
{\section{Conclusion}
	\label{sec:conclusion}}
\glspl{asic} and \glspl{fpga} are often used to hardware accelerate the convolution layers of a \gls{cnn} where up to 90\% of the computation time is consumed. This computation requires large amounts of multipliers as part of the many thousands of \gls{mac} operations needed in the convolution layer. These multipliers consume large amounts of physical and computational \gls{ic} die resources or \gls{dsp} units on a \gls{fpga}. Hardware accelerators have been proposed that reduced the amount of kernel data required by the neural network by dictionary compressing the weight values after training the network. This ``weight sharing'' reduces the bandwidth and power of the data transfers from external memory but still requires large numbers \gls{mac} units.

We reduce power and area of the \gls{cnn} accelerator by implementing \gls{pasm} in a weight-shared \gls{cnn} accelerator. \gls{pasm} re-architects the \gls{mac} to instead count the frequency of each weight and place it in a bin. The accumulated value is computed in a subsequent multiply phase, significantly reducing gate count and power consumption of the \gls{cnn}. We coded in Verilog a 16-MAC weight-shared accelerator and a 16-PAS-4-MAC weight-shared-with-\gls{pasm} accelerator and compare the logic resource requirements of a $b = 16$ bin for varying w bit widths. Gate counts are normalized to a NAND2X1 gate. For $w = 32$ bits wide the 16-PAS-4-MAC has overall 66\% fewer logic gates and consumes 70\% less total power than the 16-MAC.

To further evaluate the efficiency gains of \gls{pasm}, we implement \gls{pasm} in a weight-sharing \gls{cnn} accelerator. We compare it to a non-weight-shared accelerator and a weight-shared accelerator, targeted at a 1GHz 45nm \gls{asic}. The gate count area and power consumption for the weight-shared-with-\gls{pasm} is lower compared to the weight-shared version. For a 4-bin weight-shared-with-\gls{pasm} accelerator that accepts a $5 \times 5$ image with a $3 \times 3$ kernel and $15$ input channels and $2$ output channels, an \gls{asic} implementation of \gls{pasm} saves 48\% NAND2 gates and 53.2\% power when compared to its weight-shared counterpart, with only a 12\% increase in latency.

We show that the weight-shared-with-\gls{pasm} accelerator can be implemented in a resource-constrained \gls{fpga}. For an accelerator with the same dimensions as the \gls{asic} version implemented on the \gls{fpga} to run at 200MHz, \gls{pasm} uses 99\% fewer \glspl{dsp} and 28\% fewer \glspl{bram} compared with the weight-shared design. For 16-bin \gls{pasm}, \gls{pasm} uses 18\% less total power when compared with its weight-shared version.

Even if INT8 approximations are desired for the weight data, an 8-bit wide, 4-bin \gls{pasm} accelerator running at 200MHz on the \gls{fpga} uses 99\% fewer \glspl{dsp} and 28\% fewer \glspl{bram} compared with the weight-shared design. An INT8 operation \gls{pasm} also uses 47\% less total power when compared with its INT8 weight-shared version.

Quantization and weight-sharing neural networks are active research areas, particularly for  reducing DRAM bus bandwidth usage and in applications such as \glspl{rnn} and \gls{lstm} networks. Weight sharing allows for implementation of a \gls{cnn} in small, low power embedded systems as less \gls{ram} is required to store the weight values. Weight sharing also offers a more rapid way of implementing an inference network on a small memory embedded device without the large training phase required of, say a Binary Neural Network (BNN). Weight sharing is used in other types of networks such as regional-\glspl{cnn}, \glspl{rnn} and \glspl{lstm} so \gls{pasm} may be a good fit there too. Wherever the number of shared weights is sufficiently small, PASM units may be an attractive alternative to a conventional weight-sharing MAC unit.

\vspace{0.3cm}
\begin{acks}
	This research is supported by Science Foundation Ireland, Project 12/IA/1381. We also extend our thanks and appreciation to the Institute of Technology Carlow, Carlow, Ireland for their support.
\end{acks}

\bibliographystyle{ACM-Reference-Format}
\bibliography{bibliography}